\documentclass[%
 reprint,
superscriptaddress,
amsmath,
amssymb,
aps,
pra,
longbibliography,
]{revtex4-2}

\usepackage{graphicx}
\usepackage{dcolumn}
\usepackage{bm}
\usepackage{xcolor}
\usepackage[T1]{fontenc}
\usepackage{times}
\usepackage{amsmath,amsfonts,amssymb,mathtools}
\usepackage{hyperref}
\usepackage{bbold}
\usepackage{tensor}
\usepackage{multirow}
\usepackage{diagbox}

\begin{document}
\preprint{APS/123-QED}

\title{Optomechanically induced transparency in Four-wave mixing atomic ensemble assisted Laguerre-Gaussian vortex cavity system}
\author{Yue-Tong Hao}
 \affiliation{Center for Quantum Sciences and School of Physics, Northeast Normal University, Changchun 130024, P. R. China.}
 \author{Yi-Mou Liu}
  \email{liuym605@nenu.edu.cn}
  \affiliation{Center for Quantum Sciences and School of Physics, Northeast Normal University, Changchun 130024, P. R. China.}
\date{\today}

\begin{abstract}
We investigate the steady-state optical response of a Laguerre-Gaussian vortex cavity system integrated with cold atoms featuring a double-$\Lambda$ energy level structure. Within this hybrid system, the atoms are driven by a cavity mode and three coherent vortex beams, each carrying independent orbital angular momentum. We first check the steady-state output spectrum of the hybrid system in the passive/active case (without/with external cavity driving).~Our findings reveal that the optomechanically induced transparency spectrum is modulated by the orbital angular momentum difference $(\Delta\ell\hbar)$ from the atomic component throughout the four-wave mixing process. The resulting loop phase ($\Delta\ell\theta$) can achieve a switching effect on the absorption and gain behavior of the hybrid system for the probe beam. Additionally, the group delay, indicative of fast/slow light phenomena, is also tuned by $\Delta\ell$. We further display how the atomic orbital angular momentum modulates the periodicity of the output spot pattern in the hybrid system.~This research provides valuable insights into the modulation of optical responses in Laguerre-Gaussian vortex cavity systems.
\end{abstract}
\maketitle

\section{\label{sec:level1}Introduction}
Unlike conventional Gaussian beams, vortex beams characterized by a phase factor $e^{i l\theta}$ carry the orbital angular momentum of $l\hbar$. Here, $\theta$ represents the azimuthal angle, and $l$ is the integer winding number of the vortex, which can take any integer value \cite{OAM-fundamentals-1, OAM-fundamentals-2}. Orbital angular momentum-carrying (OAM) beams can be generated using various techniques, such as spatial light modulators (SLMs) \cite{OAM-preparation-1.1, OAM-preparation-1.2, OAM-preparation-1.3}, spiral phase plates (SPP) \cite{OAM-preparation-2.1, OAM-preparation-2.2}, Q-plat \cite{OAM-preparation-3.1}, meta-Q-plat \cite{OAM-preparation-4.1}, and liquid-crystal Damman vortex grating \cite{OAM-preparation-5.1}.
When interacting with the matter, their specific properties give rise to a variety of interesting phenomena, including vortex slow light \cite{OAM-vortex-slow-light-1, OAM-vortex-slow-light-2}, the entanglement of photons in OAM states \cite{OAM-entanglement}, transfer of optical vortices \cite{OAM-transfer-1, OAM-transfer-2, OAM-transfer-3, OAM-transfer-4, OAM-transfer-5, OAM-transfer-6, OAM-transfer-7, OAM-transfer-8, OAM-transfer-9}, light-induced torque \cite{OAM-torque-1, OAM-torque-2}, storage and retrieval of OAM-carrying beams \cite{OAM-storage-1, OAM-storage-2}, and spatially dependent electromagnetically induced transparency (EIT) \cite{OAM-EIT-1, OAM-EIT-2, OAM-EIT-3}.
These approaches offer significant advantages in terms of all-optical control and reconfigurability.
In addition, vortex beams are of interest in optical tweezers \cite{OAM-tweezer-1, OAM-tweezer-2}, imaging and microscopy \cite{OAM-microscopy-1, OAM-microscopy-2, OAM-microscopy-3}, sensing technologies for detecting molecules and nanostructures \cite{OAM-sensing-technologies-1, OAM-sensing-technologies-2}, and optical communications \cite{OAM-communications-1, OAM-communications-2, OAM-communications-3, OAM-communications-4, OAM-communications-5}.

Analogous to EIT \cite{EIT-fundamentals}, optomechanically induced transparency (OMIT) occurs when probe field photons interfere with up-converted sideband photons from the pump field via anti-Stokes processes, creating a transparency window in the resonance region. This concept was theoretically proposed by G. S. Agarwal and Huang, and experimentally demonstrated by Weis \textit{et al.}~in 2010 \cite{EIT-theory, EIT-experiments}. As an important optical phenomenon in optomechanical systems, it has been applied to many fields, such as fast and slow light \cite{EIT-fast-and-slow-light-1, EIT-fast-and-slow-light-2, EIT-fast-and-slow-light-3, EIT-fast-and-slow-light-4}, optical storage \cite{EIT-storage-1, EIT-storage-2, EIT-storage-3}, and optical switch \cite{EIT-switch-1, EIT-switch-2}. In 2007, Bhattacharya \textit{et al.} proposed a Laguerre Gaussian (L-G) cavity with Laguerre Gaussian modes \cite{LG-cavity}. 
Distinct from the Fabry-Perot (F-P) cavity, it can use the rotational degrees of freedom to couple the optical field and exchange OAM with a spiral phase element. 
This provides a promising platform for multimode macroscopic quantum phenomena.
Currently, investigations into the L-G cavity encompass a range of topics, including quantum entanglement \cite{LG-cavity-entanglement, LG-cavity-entanglement-1, LG-cavity-entanglement-2, LG-cavity-entanglement-3, LG-cavity-entanglement-4}, mechanical cooling \cite{LG-cavity-cooling-1}, and OMIT \cite{LG-cavity-OMIT-1, LG-cavity-OMIT-2, LG-cavity-OMIT-3, LG-cavity-OMIT-4}. The application of OMIT in L-G cavity system provides new tools for parameter measurement, including OAM. While OAM adds a new modulation parameter to the optomechanical system, its effect on the steady-state optomechanical response is limited. This influence is mainly seen in the unchanged rotation-optical coupling strength $g_{\phi} = cl/L$, which remains constant after preparation in L-G cavities. In general cavity optomechanical systems, the easily encoded OAM information is lost during the modulus operation, resulting in both a waste of quantum resources and a limitation on potential applications. Therefore, effectively harnessing the modulation effect of OAM in L-G cavity systems is a key issue that warrants further exploration.

In contrast to these limitations, four-wave mixing (FWM) has emerged as a versatile and powerful tool for manipulating and applying optical OAM. FWM enables OAM-based quantum operations, such as CNOT gates with L-G beams, in atomic vapors \cite{LG-FWM1}, and facilitates coherent optical vortex transfer via backward FWM in double-$\Lambda$ systems \cite{LG-FWM2}. Beyond atomic platforms, FWM efficiency can be significantly enhanced through Coulomb-mediated interactions in hybrid cavity optomechanics \cite{LG-FWM3}.~Crucially, solid-state implementations utilizing EIT demonstrate slow-light FWM in rare-earth crystals, permitting direct arithmetic operations (addition, subtraction, cancellation) of OAM topological charges without storage \cite{LG-FWM4}.~This established capability of FWM to robustly control OAM across platforms—from gaseous to solid-state quantum memories—highlights its significance for advanced photonic information processing.

In this paper, we place ultracold atoms with double-$\Lambda$ type four-level structure in a L-G cavity. Utilizing four-wave mixing processes (FWM), we try to propose a scheme to transfer orbital angular momentum from the coherent laser beams (driving atoms) to the cavity mode, with the goal of modulating the system's optical response using OAM. By examining two distinct FWM coupling cases, we find that the loop OAM, denoted as $\Delta \ell\hbar$, can switch the system's output properties between absorption ($\Delta \ell = 0$) and gain ($\Delta \ell = 2$) with $\theta = \pi/2$. This OAM-induced switch also directly influences the group delay features of the output beam, modulating both slow and fast light conditions. Further, it is explored how the output spot pattern is affected by varying the detuning between the driving frequency, the cavity mode frequency, and the probe frequency, revealing the relationship between the periodicity of the spot petals and these parameters. 

This work is organized through the following: Sec.~\ref{SecII}, where we describe the background model, and Sec.~\ref{SecIII}, where we discuss the steady-state optical response of the hybrid system, including OMIT spectra, conversion of fast/low light, optical vortices in the output field, and some error analysis. We summarize, at last, our conclusions in Sec.~\ref{SecIV}.

\section{Model and equations}\label{SecII}
\begin{figure}[ptb]
  \centering
  \includegraphics[width=0.48\textwidth]{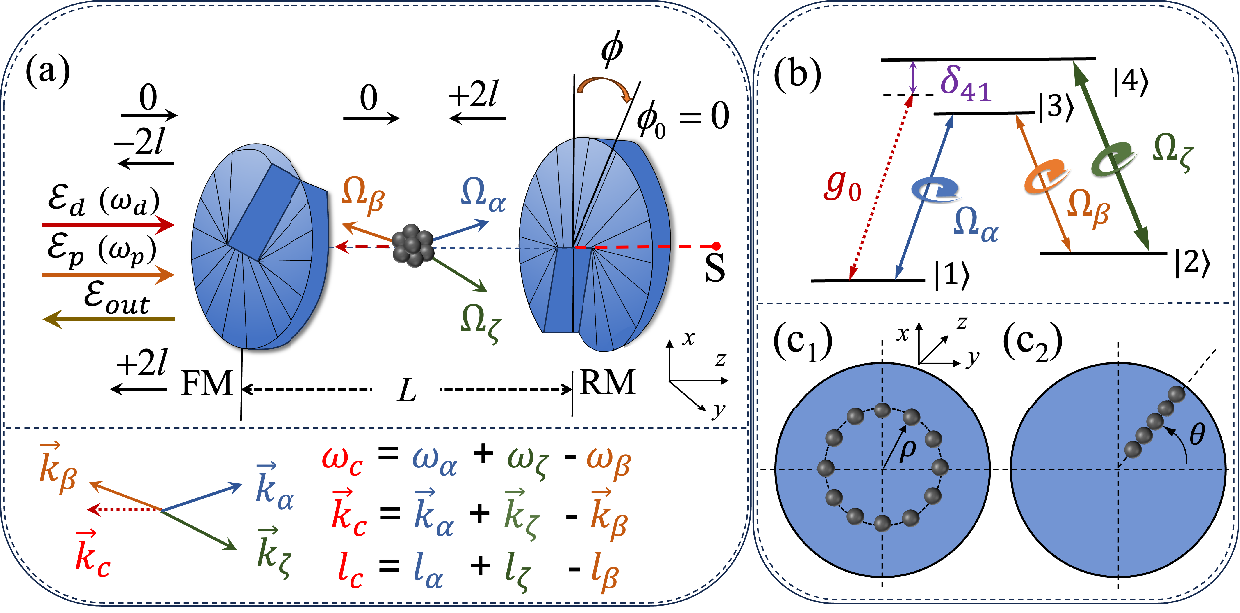}
  \caption{Schematic diagram of the system: (a) An optomechanical cavity with a rigidly fixed mirror (FM) and a rotational one (RM), the cavity mode is the L-G mode. The equilibrium position of the rotational mirror is $\phi_0$. 
  Inside the cavity, a double-$\Lambda$ four-level type atomic ensemble is driven by three coherent laser fields, and the energy level structure is shown in panel (b). The directions of the three driving fields are shown at the bottom of panel (a), and their wave vectors satisfy $\vec k_c+\vec k_{\beta}=\vec k_{\alpha}+\vec k_{\zeta}$. The topological charges ($\ell_{\mu}$, $\mu\in\{\alpha,\beta,\zeta\}$) on the vortex beams have been indicated. Two distinct spatial configurations of atoms within the cavity. Specifically, panel (c$_1$) shows atoms distributed centrosymmetrically around the optical axis $z$, while panel (c$_2$) depicts atoms distributed non-uniformly along a radial direction at a fixed angle $\theta$ to the $y$-axis.}
  \label{fig1}
\end{figure}

A Laguerre-Gaussian cavity consists of two spiral phase elements: one is fixed in place, while the other is mounted on a support structure, labeled as $S$, which enables it to rotate around the cavity axis. The fixed mirror, labeled FM in Fig.~\ref{fig1}(a), has an inversion-symmetric structure, which ensures that any beam transmitted through it retains its topological charge, while a beam reflected from FM loses a charge of $2l$. The rear rotating mirror (RM), which is perfectly reflective, adds a topological charge of $2l$ to the reflected beam. Consequently, within the L-G cavity, the orbital angular momentum of the beam propagating to the right is always $0$, while that of the beam moving to the left is consistently $2l\hbar$, under the normal Gaussian mode driving. 

As illustrated in Fig.~\ref{fig1}(a), the L-G cavity is driven by a normal Gaussian pump field (frequency $\omega_d$), and it is probed by a weak field with frequency $\omega_p$. Inside the L-G cavity, there is a double-$\Lambda$ type four-level atomic ensemble, consisting of two ground states $|1\rangle$ and $|2\rangle$ and two excited states $|3\rangle$ and $|4\rangle$, as depicted in Fig.~\ref{fig1}(b). The transition $|1\rangle\leftrightarrow|4\rangle$ is probed by the quantum cavity mode with coupling strength $g_0=\wp_{14}\sqrt{\omega_c/(2\hbar\varepsilon_0 V)}$ for single atom, where $\varepsilon_0$ ($V$, $\wp_{14}$) is the permittivity of free space (the cavity mode volume, the transition dipole moment). Meanwhile, the transitions $|1\rangle\leftrightarrow|3\rangle$, $|2\rangle\leftrightarrow|3\rangle$, and $|2\rangle\leftrightarrow|4\rangle$ are driven by classical coherent fields with half Rabi frequencies $\Omega_{\mu}=\wp_{\nu}E_{\mu}/2\hbar$ ($\mu\in\{\alpha,\beta,\zeta\}$ and $\nu\in\{13,23,24\}$), corresponding to the respective frequencies $\omega_{\alpha,\beta,\zeta}$.

The Hamiltonian of the system takes the form $H=H_{0} + H_{af} + H_{om}$, with
\begin{align}\label{Eq_H}
H_{0} & =\hbar\sum_{j=1}^{N}\sum_{q=1}^{4}\omega_{q}\hat{\sigma}_{qq}^{(j)}+\hbar\omega_{c}\hat{a}^{\dagger} \hat{a}+\frac{L_{z}^2}{2 I}+\frac{1}{2}I\omega_{\phi}^2\phi^2, \nonumber\\
  H_{af} & = -\hbar\sum_{j=1}^{N}[g(\mathbf{r}_j)  \hat{a} \hat{\sigma}_{41}^{(j)} + \Omega_{\alpha}(\mathbf{r}_j)\hat{\sigma}_{31}^{(j)}e^{-\mathrm{i}\omega_{\alpha}t}\\
  & +\Omega_{\beta}(\mathbf{r}_j)\hat{\sigma}_{32}^{(j)}e^{-\mathrm{i}\omega_{\beta}t}
  +\Omega_{\zeta}(\mathbf{r}_j)\hat{\sigma}_{42}^{(j)}e^{-\mathrm{i}\omega_{\zeta}t}]+ h.c., \nonumber\\         
H_{om} & = -{\hbar  g_{\phi}\hat{a}^\dagger \hat{a}\phi}+\mathrm{i}\hbar(\mathcal{E}_d \hat{a}^\dagger e^{-\mathrm{i}\omega_{d}t}+\mathcal{E}_p \hat{a}^\dagger e^{-\mathrm{i}\omega_{p}t}) +h.c..\nonumber 
\end{align}
Here $H_{0}$ represents the free Hamiltonian including atomic ensemble with frequencies $\omega_{q}$ ($q\in\{1,2,3,4\}$ and $\omega_{1}=0$), the cavity mode (frequency $\omega_c$) and the RM mirror (frequency $\omega_{\phi}$). $\hat{\sigma}_{pq}^{(j)}=|p\rangle_{jj}\langle q|$ is defined as the projection ($p=q$) or transition ($p\neq q$) operator of the $j$th atom  with $N$ denoting the atomic number; 
$\hat{a}$ $(\hat{a}^\dagger)$ denotes the annihilation (creation) operator of the cavity field; $I=MR^2/2$ gives the moment of inertia of the RM mirror about the cavity axis, where $M$ and $R$ are the mass and radius of the mirror, respectively; $L_{z}$ ($\omega_{\phi}$, $\phi$) is the angular momentum (the angular rotation frequency, the angular displacement) with satisfying the commutation relation $[L_{z},\phi]=-\mathrm{i}\hbar\delta$. 
And therefore, $H_{af}$ reads the atom-field and atom-cavity interactions.~$H_{om}$ contains the optomechanical interaction with the coupling strength $g_{\phi}=cl /L$ ($c$ the vacuum speed of light, $l$ the OAM quantum number of cavity mode, $L$ the cavity length) and those between the cavity and the two external fields with amplitudes $\mathcal{E}_{d,p}=\sqrt{2P_{d,p}\kappa/\hbar\omega_{d,p}}$, where $P_d$ $(P_p)$ is the power of the driving (probe) field.~Here, we assume that the cavity mode is a L-G mode,
\begin{align}
 \hat{E}(\mathbf{r},t)=\mathrm{i}\sqrt{\frac{\hbar\omega_c}{2\varepsilon_0V}}[\hat{a}\cdot\mathcal{U}(\mathbf{r})e^{-\mathrm{i}\omega_c t}-\hat{a}^{\dagger}\mathcal{U}^{*}(\mathbf{r})e^{\mathrm{i}\omega_c t}],  
\end{align}
where $\mathcal{U}(\mathbf{r})\propto (\frac{\sqrt{2}\rho}{w})^{|l|}e^{-\frac{\rho^2}{w^2}}e^{\mathrm{i}l\theta}$ is the spatial distribution function of the L-G mode, with $\rho$ ($\theta$) being the radial coordinate (azimuthal angle) in cylindrical coordinates with OAM $l\hbar$. Therefore, the coupling strength between atoms inside the cavity and the cavity mode also exhibits a spatial distribution, $g(\mathbf{r})=g_0 \cdot \mathcal{U}(\mathbf{r})$. Similarly, the three coherent fields incident on the atomic ensemble are L-G beams carrying different OAM ($\ell_{\mu}\hbar$). In this case, the Rabi frequency of the three fields are characterized as
\begin{align}\label{Eq_|omega|}   \Omega_{\mu}=\mathcal{E}_\mu(\frac{\sqrt{2}\rho}{w})^{|\ell |}e^{-\frac{\rho^2}{w^2}}e^{\mathrm{i}\ell _{\mu} \theta}, (\mu=\alpha, \beta, \zeta)
\end{align}
where $w$ represents the beam waist, and $\mathcal{E}_\mu$ is the strength of the vortex beam. The FWM process shown in Fig.~\ref{fig1} generates photons at the cavity mode frequency ($\omega_{c}=\omega_{\mathrm{FWM}}=\omega_{\alpha}-\omega_{\beta}+\omega_{\zeta}$), with wavevector $\vec k_c=\vec k_{\alpha}-\vec k_{\beta}+\vec k_{\zeta}$ and carrying OAM $\Delta\ell\hbar$ ($\Delta\ell=\ell_{\alpha}-\ell_{\beta}+\ell_{\zeta}$).

With low excitation and large number approximation of atomic ensemble, the dynamics of the system can be described by the following Heisenberg-Langevin equations:
\begin{align}\label{Eq_rotating}
  \dot\phi & = L_z/I, \nonumber\\
  \dot L_z & = -\gamma_\phi L_z-I\omega_\phi^2\phi+\hbar g_{\phi}\hat{a}^\dagger \hat{a}-\sqrt{2\gamma_\phi}\xi(t), \nonumber\\
  \dot{\hat{a}} & = -[\kappa+\mathrm{i}\Delta_c-\mathrm{i} g_{\phi}\phi]\hat{a}+\mathrm{i}\sum_j g(\mathbf{r}_j)\varrho_{14}^{(j)}
  +\mathcal{E}_d \nonumber\\ 
  &+\mathcal{E}_pe^{-\mathrm{i}\delta_p t}+\sqrt{2\kappa}a^{in}(t), \\
  \dot\varrho_{14}^{(j)} & = -[\gamma_{14}+\mathrm{i}\delta_{c}]\varrho_{14}^{(j)}+\mathrm{i} g(\mathbf{r}_j) \hat{a}[\varrho_{11}^{(j)}-\varrho_{44}^{(j)}] \nonumber\\
  & +\mathrm{i}\Omega_{\zeta}(\mathbf{r}_j)\varrho_{12}^{(j)}-\mathrm{i}\Omega_{\alpha}(\mathbf{r}_j)\varrho_{34}^{(j)}+\sqrt{2\gamma_{14}}f_{14}(t), \nonumber
\end{align}
after the following frame rotation of the above equations: 
$\hat{a}=\hat{a} e^{-\mathrm{i}\omega_d t}$, $\varrho_{14}=\hat{\sigma}_{14}e^{-\mathrm{i}\omega_d t}$, $\varrho_{13}=\hat{\sigma}_{13}e^{-\mathrm{i}\omega_{\alpha}t}$, $\varrho_{23}=\hat{\sigma}_{23 }e^{-\mathrm{i}\omega_{\beta}t}$, $\varrho_{24}=\hat{\sigma}_{24}e^{-\mathrm{i}\omega_{\zeta}t}$, $\varrho_{12}=\hat{\sigma}_{12}e^{-\mathrm{i}(\omega_d-\omega_{\zeta})t}$, $\varrho_{34}=\hat{\sigma}_{34}e^{-\mathrm{i}(\omega_d-\omega_{\alpha})t}$.~Here $\Delta_{c}=\omega_c-\omega_d$ (denoting the detuning between the cavity mode and the driving field), $\delta_{p}=\omega_p-\omega_d$ (denoting the detuning between the probe and the driving field), $\delta_{c}=\omega_{4}-\omega_c$ (denoting the detuning between the cavity mode and the corresponding energy level of the atom), 
$\delta_{31}=\omega_{3}-\omega_{\alpha}$, $\delta_{21}=\delta_{c}-\delta_{42}=\omega_{2}-\omega_d+\omega_{\zeta}$. 
Assuming all the fields couple to the atomic ensemble resonantly, we have $\delta_{31}=\delta_{32}=\delta_{42}=0$. In addition, we phenomenologically add the mirror damping rate (the cavity decay rate, the atomic dephasing rates) as $\gamma_\phi$ ($\kappa$, $\gamma_{pq}$); and the term $\xi$ ($a^{in}$, $f_{pq}$) is also introduced to represent the quantum fluctuations of the rotating mirror (the cavity field, and the external field driving the atoms).

Considering the quantum fluctuations of each operator $\hat{O}\in\{\phi, L_z,\hat{a}\}$ in the steady state, we linearize around the steady-state value $O_s$ by adding the fluctuation operator $\delta O$, such that $\hat{O}=O_s+\delta O$. 
Then we can obtain the following steady-state solutions
\begin{align}\label{Eq_state_solutions}
  \phi_{s}  & = \frac{\hbar  g_\phi|a_s|^2}{I\omega_\phi^2}, \nonumber\\
  a_{s} & = \frac{\mathcal{E}_d+\mathrm{i}\sum_jg(\mathbf{r}_j)B_f(\mathbf{r}_j)}{\kappa+\mathrm{i}\bar\Delta-\mathrm{i}\sum_jg(\mathbf{r}_j)A_f(\mathbf{r}_j)}, \nonumber\\
  \varrho_{14,s}^{(j)}  & = A_f(\mathbf{r}_j) a_s+B_f(\mathbf{r}_j), (f=0,1) 
\end{align}
where $\bar\Delta = \Delta_c-g_{\phi}\phi_s$ denotes the effective detuning between the cavity field modes and the driving field when the rotating mirror deviates from its equilibrium position. Equation \eqref{Eq_state_solutions} shows that the FWM atoms introduce two effects to the system:
first, the cavity decay rate increases by $\text{Im}[\sum_jg(\mathbf{r}_j)A_f(\mathbf{r}_j)]$; 
second, the cavity frequency is effectively shifted by $\text{Re}[\sum_jg(\mathbf{r}_j)A_f(\mathbf{r}_j)]$.
It is worth noting that $\varrho_{14,s}$ contains two terms, the first of which represents a linear term $A_f a_s$ and the second reads the four-wave mixing term $B_f$ \eqref{Eq_liouville_af}. In this study, two conditions are considered: one in which $\Omega_\alpha$ is a weak field with 
\begin{align}\label{Eq5}
  A_0(\mathbf{r}_j) & =\frac{g(\mathbf{r}_j)[\delta_{c}\gamma_e-\mathrm{i}|\Omega_\beta(\mathbf{r}_j)|^2]}{DD_0(\mathbf{r}_j)} , \nonumber\\
  B_0(\mathbf{r}_j) & =\frac{\mathrm{i}\Omega_\alpha(\mathbf{r}_j) \Omega_{\beta}^{*}(\mathbf{r}_j) \Omega_{\zeta}(\mathbf{r}_j)}{DD_0(\mathbf{r}_j)}.
\end{align}
 The other case is for $\Omega_{\beta}$ as  weak field ($\Omega_\alpha=\Omega_\zeta=\Omega$) with 
\begin{align}
  A_1(\mathbf{r}_j) & =\frac{-g(\mathbf{r}_j)\delta_{c}\gamma_e(\gamma_e^2-|\Omega(\mathbf{r}_j)|^2+\mathrm{i}\delta_{c}\gamma_e/2)}{DD_1(\mathbf{r}_j)}, \nonumber\\  B_1(r,\theta) & =\frac{\Omega^2(\mathbf{r}_j)\Omega_\beta^*(\mathbf{r}_j)\gamma_e(\delta_{c}-2\mathrm{i}\gamma_e)}{DD_1(\mathbf{r}_j)}. 
\end{align}
Moreover, the corresponding denominators read 
\begin{align} 
  DD_0 & =\delta_{c}^2\gamma_e-(|\Omega_{\beta}|^2+|\Omega_{\zeta}|^2)\gamma_e-\mathrm{i}\delta_{c}(|\Omega_{\beta}|^2+\gamma_e^2),\\
DD_1 & =(\mathrm{i}\delta_{c}+\gamma_e)(4|\Omega|^2-\delta_{c}^2+2\mathrm{i}\delta_{c}\gamma_e)(2|\Omega|^2+\gamma_e^2),\nonumber
\end{align}
here, with $\Omega_{\mu}$ ($\mu\in\{\alpha,\beta,\zeta\}$) exhbiting a apatial dependence on $\mathbf{r}$.

In general, the mean value of each dynamic variable considered here is much larger than its fluctuation, i.e.~$|O_s|\gg\delta O$. 
In this condition, one can safely neglect the nonlinear terms, e.g.~$\delta a\delta a^\dagger$. 
Similarly, we can get the following linearized quantum Langevin equations \cite{EIT-switch-2,LG-cavity}: 
\begin{align}\label{Eq_fluctuate}
  \delta\dot\phi & = \delta L_z/I, \nonumber\\
  \delta\dot L_z & = -I\omega_\phi^2\delta\phi+\hbar g_{\phi}(a_s^*\delta a+a_s\delta a^\dagger)-\gamma_\phi \delta L_z, \nonumber\\
  \delta\dot{a} & = -(\kappa+\mathrm{i}\Delta_c)\delta a+\mathcal{E}_pe^{-\mathrm{i}\delta_p t}+\mathrm{i} \sum_j g(\mathbf{r}_j)\delta\varrho_{14}^{(j)}\nonumber\\
   & +\mathrm{i}g_{\phi}[\phi_s \delta a+a_s \delta \phi], \nonumber\\ 
  \delta\dot{\varrho}_{14}^{(j)} & = A_f(\mathbf{r}_j) \delta a+ B_f(\mathbf{r}_j).
\end{align}

\section{Steady-state Optical Response of the Output Field}\label{SecIII}
In this section, we focus on the steady-state optical response, including absorptive and dispersive characteristics, of the output field. We also investigate the correlation between these characteristics and the orbital angular momentum imparted by the driving field of the cavity mode and the coupled atoms, as well as the coupling strength of the cavity field. The parameters employed are sourced from the referenced studies \cite{LG-cavity,LG-cavity-entanglement}, including the cavity length $L = 0.2$ mm, the wavelength of the driving field $\lambda = 810$ nm, the cavity radius $R = 10$ $\mu$m, the cavity mass $M = 100$ ng, the mirror's angular frequency $\omega_\phi = 2\pi \times 35$ MHz, the mirror's decay rate $\gamma_\phi = 2\pi \times 140$ Hz, and the cavity's decay rate $\kappa = 2\pi \times 8$ MHz. The Rabi frequencies for the three laser beams are set at $\Omega_\alpha^{(\mathrm{max})} = 0.02 \times 2\pi$ MHz, $\Omega_\beta^{(\mathrm{max})} = 10 \times 2\pi$ MHz, and $\Omega_\zeta^{(\mathrm{max})} = 10 \times 2\pi$ MHz, respectively.

To examine the dynamic stability of the system, we further rewrite the Langevin equations by $\dot u(t) = Mu(t)+v(t)+\eta(t)$, which are expressed in terms of the quadrature operators 
$u(t)=(\delta\phi,\delta L_z,\delta U,\delta V)^T$ with $U=\frac{1}{\sqrt{2}}(\delta a+\delta a^\dagger)$ and $V=\frac{1}{\sqrt{2}\mathrm{i}}(\delta a-\delta a^\dagger)$ being the amplitude and phase quadratures of the cavity field.~We also set $v(t)=(0,0,\frac{1}{\sqrt{2}}(\varepsilon_pe^{-\mathrm{i}\delta_pt}+\varepsilon_p^*e^{\mathrm{i}\delta_pt})-\sqrt{2}\sum_jg(\mathbf{r}_j)\text{Im}[B_f(\mathbf{r}_j)],\frac{1}{\sqrt{2}\mathrm{i}}(\varepsilon_pe^{-\mathrm{i}\delta_pt}-\varepsilon_p^*e^{\mathrm{i}\delta_pt})+\sqrt{2}\sum_jg(\mathbf{r}_j)\text{Re}[B_f(\mathbf{r}_j)])^T$ as the entirely passive input noise part originating from the probe field and the atomic coupling, and $\eta(t)=(0,-\sqrt{2\gamma_\phi}\delta\xi,\sqrt{2\kappa}\delta U^{in}$, $\sqrt{2\kappa}\delta V^{in})$ with $\delta U^{in}=\frac{1}{\sqrt{2}}(a_{in}+a_{in}^\dagger)$, $\delta V^{in}=\frac{1}{\sqrt{2}\mathrm{i}}(a_{in}-a_{in}^\dagger)$ can be regarded as the input vacuum noises. Therefore, the evolution matrix reads
\begin{align}\label{Eq_M}
  M=\begin{pmatrix} 0 & 1/I & 0 & 0 \\ -I\omega_\phi^2 & -\gamma_\phi & \sqrt{2}\hbar g_\phi a_r & \sqrt{2}\hbar g_\phi a_i \\ -\sqrt{2} g_\phi a_i & 0 & -\kappa ^{\prime} & \tilde{\Delta}_{A_f} \\ \sqrt{2} g_\phi a_r& 0& -\tilde{\Delta}_{A_f} & -\kappa^{\prime}\end{pmatrix},
\end{align}
with $a_r=\frac{a_s+a_s^*}{2}=\text{Re}[a_s]$, $a_i=\frac{a_s-a_s^*}{2\mathrm{i}}=\text{Im}[a_s]$, $\kappa^{\prime}=\kappa-\sum_j g(\mathbf{r}_j)\text{Im}[A_f (\mathbf{r}_j)]$ and $\tilde{\Delta}_{A_f}=\bar{\Delta}_c-\sum_jg(\mathbf{r}_j)\text{Re}[A_f(\mathbf{r}_j)]$.

Followed by the Routh-Hurwitz criterion \cite{R-H}, the real part of the eigenvalues of $M$ should be strictly negative such that the system is stable, which gives rise to the stability conditions below
\begin{align}\label{Eq_s}
  s_1 & = (2\kappa^{\prime}+\gamma_\phi)[\tilde{\Delta}_{A_f}^2+{\kappa^{\prime}}^2+2\gamma_\phi\kappa^{\prime}+\omega_\phi^2] \nonumber\\
       & -[\gamma_\phi \tilde{\Delta}_{A_f}^2+\gamma_\phi{\kappa^{\prime}}^2+2{\kappa^{\prime}}\omega_\phi^2]>0, \\
  s_2 & = -2\hbar g_\phi^2 \tilde{\Delta}_{A_f}(a_r^2+a_i^2)+I\omega_\phi^2[{\kappa^{\prime}}^2+\tilde{\Delta}_{A_f}^2]>0, \nonumber\\
  s_3 & = [\gamma_\phi I\tilde{\Delta}_{A_f}^2+\gamma_\phi^2I\kappa'^2+2I\kappa^{\prime}\omega_\phi^2]s_1-(2\kappa^{\prime}+\gamma_\phi)^2s_2>0.\nonumber
\end{align}
The investigation revealed that the range of parameters selected is consistent with the above inequality.

\begin{figure}[ptb]
  \includegraphics[width=0.48\textwidth]{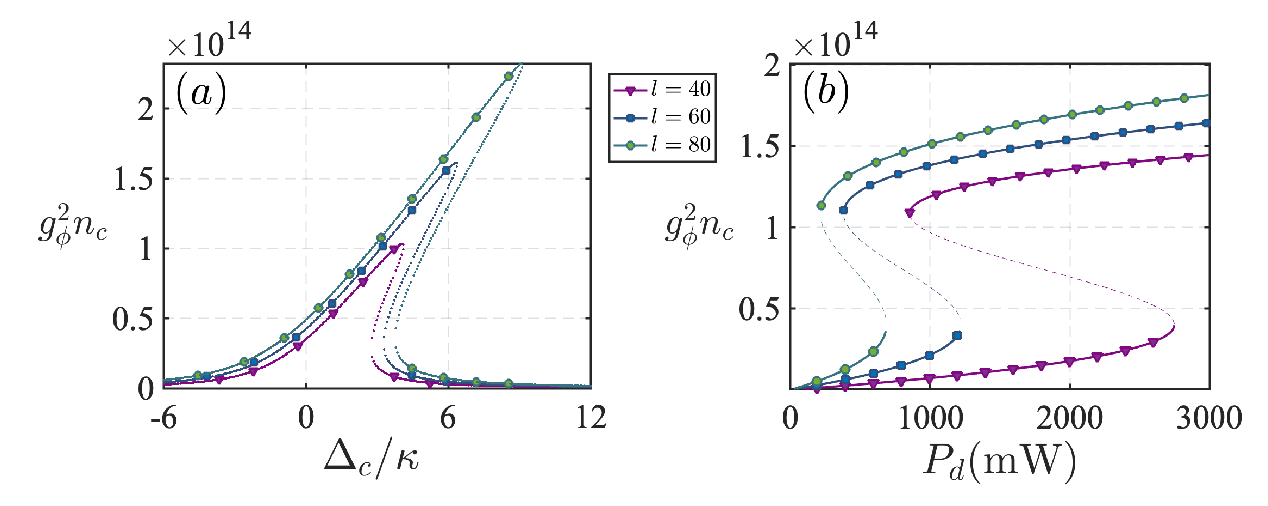}
  \caption{The normalized cavity intensity $g_\phi^2 n_c$ (a) as a function of the dimensionless detuning $\Delta_c/\kappa$ for driving power $P_d$ = 200 mW
  and (b) as a function of the driving power $P_d$ for $\Delta_c/\kappa$ = 35, with $l=40$ (magenta-triangle), 60 (blue-square), 80 (olive-circle). }
  \label{Fig_multisteady}
\end{figure}
From Eq. \eqref{Eq_state_solutions}, there is a higher-order equation for the average photon number, which predicts a possible multistability of the system and can be described by the following polynomial:
\begin{align}\label{Eq_bistable}
    & \frac{\hbar^2g_{\phi}^4}{I^2\omega_\phi^4}n_c^3-\frac{2\hbar g_{\phi}^2\tilde{\Delta}_{A_f}}{I\omega_\phi^2}n_c^2 
     +[\kappa^{\prime2}+\tilde{\Delta}_{A_f}^2]n_c\\ 
     & = \sum_jg^2(\mathbf{r}_j)|B_f(\mathbf{r}_j)|^2-2\sum_jg(\mathbf{r}_j)\mathcal{E}_d \text{Im}[B_f(\mathbf{r}_j)]+\mathcal{E}_d^2,\nonumber
\end{align}
which is a cubic equation about $n_c$ and can be multistable. Fig.~\ref{Fig_multisteady}(a) examines the effective optomechanical coupling strength, given by $|G|^2 \propto g_{\phi}^2 n_c$, where $n_c=\langle a_s^{\dagger}a_s\rangle$ is the intracavity average photon number, across various detunings $\Delta_c$ between the cavity mode and the driving field. As anticipated, when $l$ becomes sufficiently large $( g_{\phi} = cl / L $), this hybrid system exhibits multistability. With $\Delta_c = 30 \times 2\pi$ MHz, the multistable behavior is illustrated in Fig.~\ref{Fig_multisteady}(b) under different driving powers $P_d$. Eq.~\eqref{Eq_bistable} further reveals that both $\Delta_c$ and $l$ ($g_{\phi} = cl / L$) strongly influence the coefficient of $n_c$, while $\mathcal{E}_d$ (or $P_d$) primarily controls the magnitude of the constant term, particularly with weak atom-cavity coupling $\sum_jg(\mathbf{r}_j)<\kappa$. Further, due to the weak nonlinearity $B_f \ll \sum_jg(\mathbf{r}_j)$, the phase of OAM introduced by the atomic four-wave mixing process has a minimal effect on the stability of the system. Under the constraint that the driving power is limited to $ P_d < 230 \text{ mW} $, we can safely analyze the optical response of the hybrid system in the single steady-state parameter range.

\subsection{OMIT spectrum and fast/slow light conversion}
We further check the steady optical response of the hybrid system. For simplicity, we will quantize the simple harmonic rotors (RM) as phonons, represented by creation (annihilation) operators $b^{\dag}$ ($b$),
\begin{align}\label{Eq_b}
  \phi & =\sqrt{\frac{\hbar}{2I\omega_{\phi}}}\left(b^\dagger+b\right),\nonumber\\
  L_{z} & =\mathrm{i}\sqrt{\frac{\hbar I\omega_{\phi}}{2}}\left(b^\dagger-b\right).
\end{align}
Substituting the above transformations into Eq.~\eqref{Eq_fluctuate}, we express the fluctuation operator in the following form: 
\begin{align} \label{Eq_fluctuate_transformations} 
    \delta a & = \delta a_{+}\mathcal{E}_p e^{-\mathrm{i}\delta_p t}+\delta a_{-}\mathcal{E}_p^* e^{\mathrm{i}\delta_p t}, \nonumber\\
    \delta b & =\delta b_{+}\mathcal{E}_p e^{-\mathrm{i}\delta_p t}+\delta b_{-}\mathcal{E}_p^* e^{\mathrm{i}\delta_p t},\\
    \delta \varrho_{14} & = \delta \varrho_{14+}\mathcal{E}_p e^{-\mathrm{i}\delta_p t}+\delta \varrho_{14-}\mathcal{E}_p^* e^{\mathrm{i}\delta_p t} \nonumber\\
    & = [A_f \delta a_{+} + \tilde{B}_f ]\mathcal{E}_p e^{-\mathrm{i}\delta_p t}+ [A_f^*\delta a_{-} + \tilde{B}_f^* ]\mathcal{E}_p^* e^{\mathrm{i}\delta_p t}, \nonumber
\end{align}
Thus, we obtain the fluctuations in the positive- and negative-frequency components of the cavity field
\begin{align}\label{Eq_delta_a2}
  \delta a_{+} & = \frac{1-\mathrm{i} \sum_j g(\mathbf{r}_j) B_f(\mathbf{r}_j)}{\kappa-\mathrm{i}[\delta_p -\sum_j g(\mathbf{r}_j) A_f(\mathbf{r}_j)]-\frac{G^2\cdot n_c}{(\gamma_{\phi}+\mathrm{i}\delta_p)}}, \nonumber\\
  \delta a_{-} & = \frac{-\mathrm{i} \sum_j g(\mathbf{r}_j) B_f(\mathbf{r}_j)}{\kappa+\mathrm{i}[\delta_p +\sum_j g(\mathbf{r}_j) A_f(\mathbf{r}_j)]-\frac{G^2\cdot n_c}{(\gamma_{\phi}+\mathrm{i}\delta_p)}}, 
\end{align}
where $Ga_s$ is defined as the effective optomechanical interaction rate ($G=g_{\phi}\sqrt{\hbar/2I\omega_{\phi}}= \frac{cl}{L}\sqrt{\hbar/2I\omega_{\phi}}$). 
Decomposing into radial and angular components under the cylindrical coordinate system, the spatial distribution functions of all L-G modes read 
\begin{align}
  g(\mathbf{r}) & =g(\rho)e^{\mathrm{i}2l\theta}\\
  \Omega_{\mu}(\mathbf{r}) & = \Omega_{\mu}(\rho)e^{\mathrm{i}\ell_{\mu}\theta},\nonumber
\end{align}
Therefore, Eq.~\eqref{Eq_delta_a2} will transform as 
\begin{align}\label{Eq_delta_a3}
  \delta a_{+} & = \frac{1-\mathrm{i} \sum_j g(\rho_j) B_f(\rho_j)e^{\mathrm{i}\ell_{\Theta }\theta}}{\kappa-\mathrm{i}[\delta_p -\sum_j g(\rho_j) A_f(\rho_j)]-\frac{G^2\cdot n_c}{(\gamma_{\phi}+\mathrm{i}\delta_p)}}, \nonumber\\
  \delta a_{-} & = \frac{-\mathrm{i} \sum_j g(\rho_j) B_f(\rho_j)e^{\mathrm{i}\ell_{\Theta }\theta}}{\kappa+\mathrm{i}[\delta_p +\sum_j g(\rho_j) A_f(\rho_j)]-\frac{G^2\cdot n_c}{(\gamma_{\phi}+\mathrm{i}\delta_p)}}. 
\end{align}

Here, $\ell_{\Theta }\theta=(\Delta\ell+2l)\theta$ is defined as the loop phase formed by the backward-propagating cavity mode ($2l$) and three control L-G beams via FWM structure. It is crucial to emphasize that if atoms within the cavity are uniformly distributed or exhibit central symmetry along the z-axis (as shown in Fig.~\ref{fig1} c$_1$), the angular distribution becomes negligible through summation (or integration) for cases where $\ell_{\Theta}\neq 0$ with $\theta \in [0,2\pi]$. 
To demonstrate the impact of angular distribution on the system's optical response, we deliberately arrange the cavity atoms in a non-uniform configuration. One of the simplest cases (Fig.~\ref{fig1} c$_2$) features atoms distributed exclusively along the angular direction $\theta$. Taking $\theta=\pi/2$ as an example, Eq.~\eqref{Eq_delta_a3} can be further reduced to 
\begin{align}\label{Eq_delta_a4}
\delta a_{+}=\frac{1-\mathrm{i} \sum_j g(\rho_j) B_f(\rho_j)e^{\mathrm{i}\mp\Delta\ell\pi/2}}{\kappa-\mathrm{i}[\delta_p- \sum_j g(\rho_j) A_f(\rho_j)] -\frac{G^2\cdot n_c}{(\gamma_{\phi}+\mathrm{i}\delta_p)}},
\end{align}
with $e^{\mathrm{i}\ell_{\Theta}\pi/2}=e^{\mp\mathrm{i}\Delta\ell\pi/2}$ for $l=2m-1$ or $2m$ ($m\in\mathbb{Z} $).

The output field of the cavity is defined as $\mathcal{E}_{out}$, using the standard input-output relation,
\begin{align}\label{Eq_in_out_re}
    \mathcal{E}_{out}+\mathcal{E}_{p}e^{-\mathrm{i} \delta_p t}+\mathcal{E}_{d}=2\kappa a. 
\end{align}
Expanding the output field as $\mathcal{E}_{out}=\mathcal{E}_{out}^s+\mathcal{E}_{out}^{+}\mathcal{E}_{p}e^{-\mathrm{i}\delta_p t}+\mathcal{E}_{out}^{-}\mathcal{E}_{p}e^{\mathrm{i}\delta_p t}$, where $\mathcal{E}_{out}^s$ denotes the steady-state form of the output field.
Substituting it into Eq. \eqref{Eq_in_out_re}, the output transmission spectrum can be obtained as 
\begin{align}\label{Eq_out}
    \mathcal{E}_{out}^s & = 2\kappa a_{s}-\mathcal{E}_{d}, \nonumber\\
    \mathcal{E}_{out}^{+} & = 2\kappa \delta a_{+}-1, \nonumber\\
    \mathcal{E}_{out}^{-} & = 2\kappa \delta a_{-}.  
\end{align}
In order to study the response of the system to the probe field, the amplitude of the rescaled output to the probe field is
\begin{align}\label{Eq_out_T}
\mathcal{E}_T=\mathcal{E}_{out}^{+}+1=2\kappa\delta a_{+}=u_p+\mathrm{i}v_p, 
\end{align}
where $u_p$ and $v_p$ represent the absorptive and dispersive coefficients of the optomechanical system, respectively.
Moreover, the expression for the transmitted probe field is defined as $t_p=1-2\kappa\delta a_{+}=1-u_p-\mathrm{i}v_p$.

\begin{figure}[ptb]
  \includegraphics[width=0.48\textwidth]{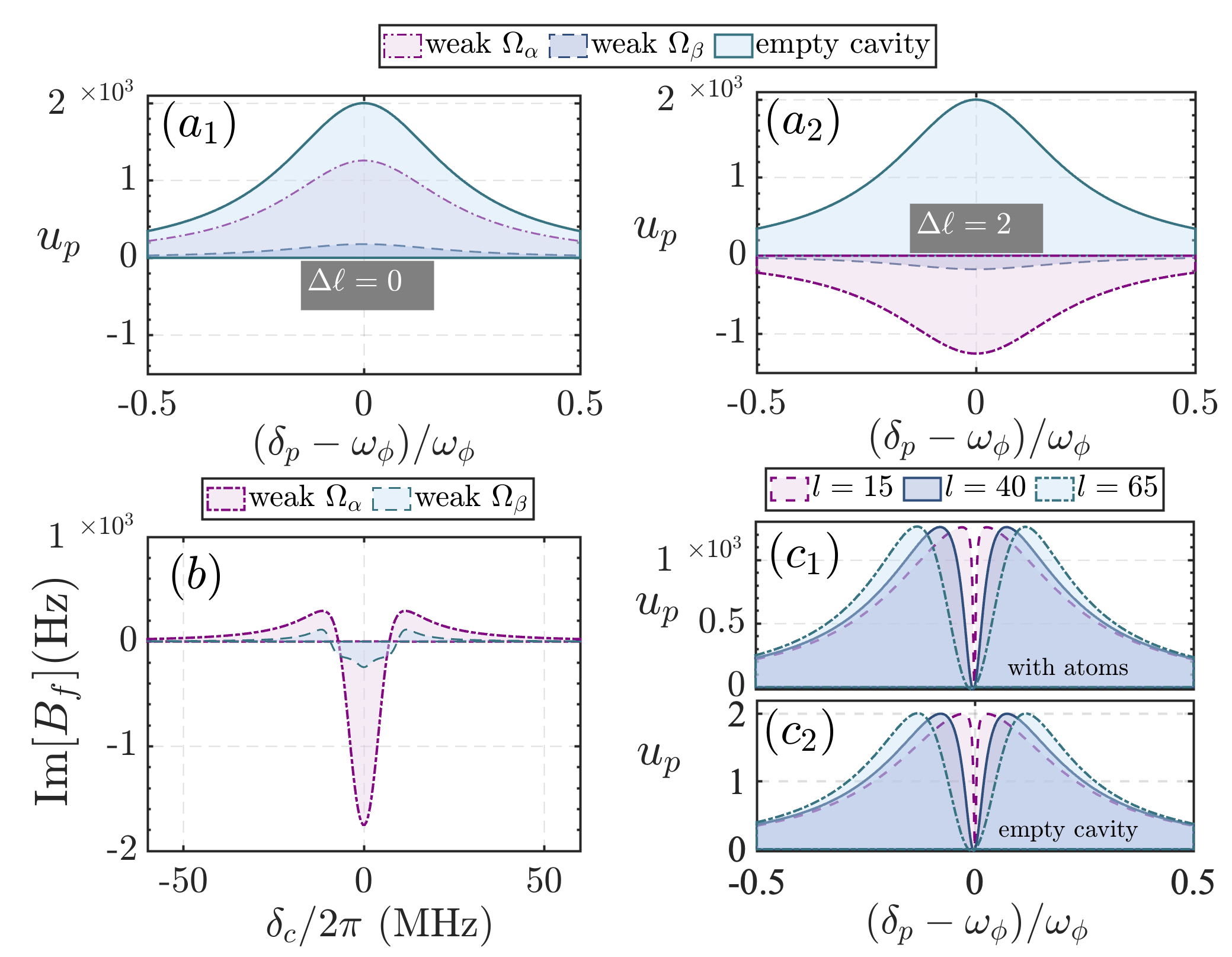}
  \caption{The absorption curves as functions of the normalized optical detuning with (a$_1$) $\Delta \ell =0$ and (a$_2$) $\Delta \ell =2$ for $\theta=\pi/2$ without cavity driving field ($P_d=0$ mW) in conditions of weak field $\Omega_{\alpha}^{(\mathrm{max})}=0.02\times 2\pi$ MHz, or weak field $\Omega_{\beta}^{(\mathrm{max})}=0.02\times 2\pi$ MHz, and empty cavity, respectively.
  Comparison of nonlinear effects at weak fields (b) for $\Omega_{\alpha}$ and $\Omega_{\beta}$, respectively.  The absorption curves (c$_1$) as a function of the normalized optical detuning with different $l $ in the case of $\Delta \ell = 0$, with Panel (c$_2$) displaying the  empty cavity cases  for comparison.~Other parameters take the values of $\sum_jg(\rho_j)= 10^{-2}\gamma_e$, $P_d = 200$ mW.}
  \label{Fig_nodriving}
\end{figure}

Initially, it is discussed that the absorption spectrum $v_p(\omega)$ of the hybrid system in the absence of any extra-cavity driving (\emph{Passive case}), as depicted in Fig.~\ref{Fig_nodriving}(a$_1$) and (a$_2$). The light magenta-dotted curves and blue-dashed curves correspond to different weak-field coupling conditions for the atoms, namely $\Omega_\alpha=0.02\times 2\pi$ MHz and $\Omega_\beta=0.02\times 2\pi$ MHz, respectively. The olive-solid curves, on the other hand, represent those of the empty cavity for comparison (Lorentz-type curve). The presence of atoms modifies the absorption spectrum of the hybrid system (Laguerre-Gaussian rotational-cavity), with the effective damping rate of the cavity $\kappa_{\mathrm{eff}}=\kappa-\sum_j\frac{g^{2}(\rho_j)|\Omega_\beta(\rho_j)|^2}{\gamma_e[|\Omega_\beta(\rho_j)|^2+|\Omega_\zeta(\rho_j)|^2]}$ [\ref{A3}]. Obviously, the phase information of the laser fields impinging on the atoms, influenced by OAM ($ \Delta\ell\hbar$), produces distinct impacts on the absorption profile of the output field. This manifests as an absorption (gain) when $\Delta\ell =0$ ($\Delta\ell =2$), with $\theta =\pi/2$ and $l=2m$. Moreover, the influence is more pronounced when $\Omega_\alpha$ represents a weak field ($\Omega_{\alpha}\ll\Omega_{\beta}\simeq\Omega_{\zeta}$). With $\delta_c=0$, we can obtain a stronger atomic nonlinear absorption $B_0\simeq\frac{-\mathrm{i}\Omega_{\alpha}}{2\gamma_e}$ than those  ($B_1\simeq\frac{-\mathrm{i}\Omega_{\beta}^{*}}{2\gamma_e+\frac{4\Omega^2}{\gamma_e}}$) for the case $\Omega_{\beta}\ll\Omega_{\alpha}\simeq\Omega_{\zeta}=\Omega$. It is also depicted in Fig.~\ref{Fig_nodriving}(b), under the assumption that both of the other two strong coherent coupling beams have a Rabi frequency of $10$ MHz.

\begin{figure}[ptb] 
\includegraphics[width=0.48\textwidth]{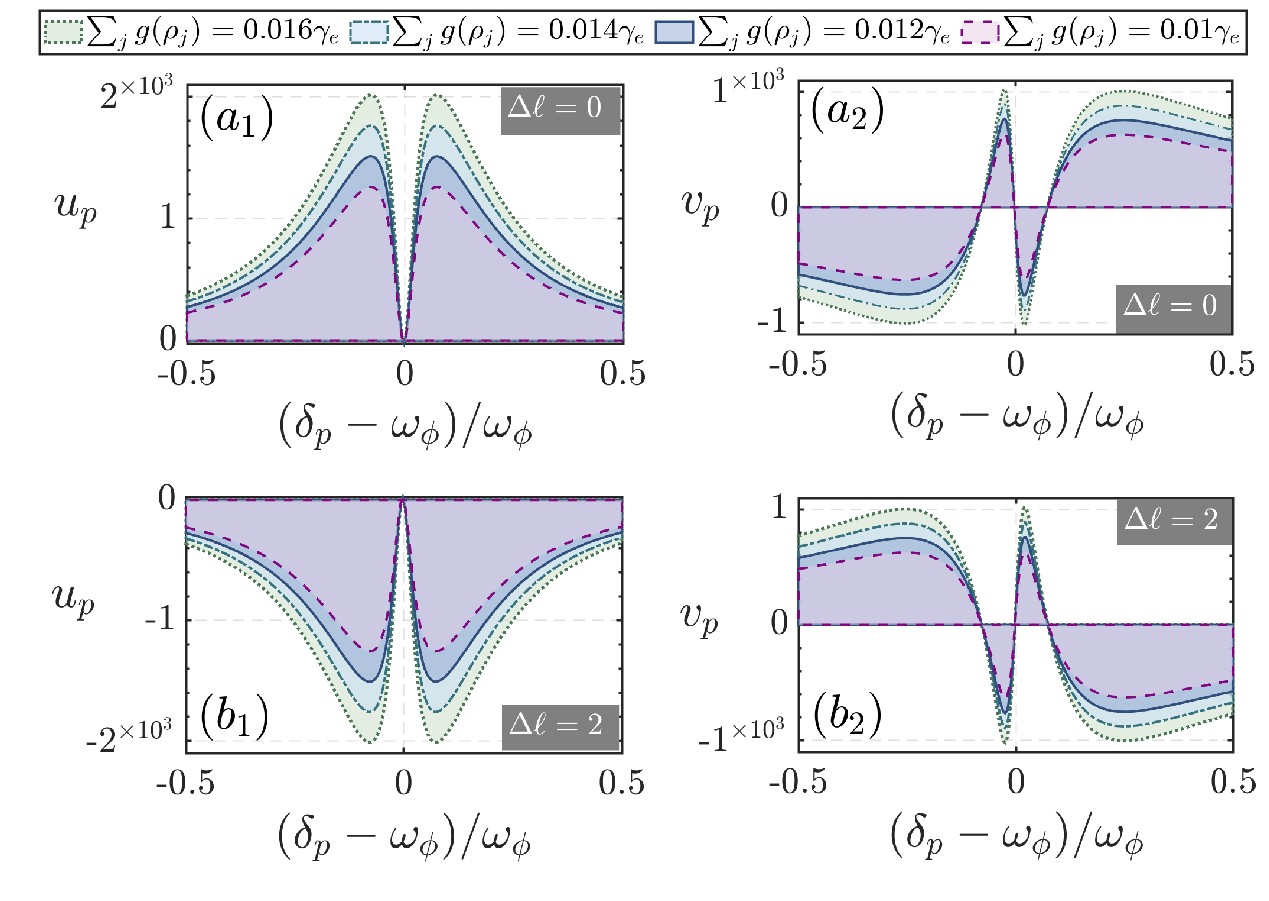}
\centering
\caption{The absorption curves (a$_1$) and dispersion curves (a$_2$) as functions of the normalized optical detuning with weak $\Omega_{\alpha}^{(\mathrm{max})}=0.02\times 2\pi$ MHz in the case of $l=40$, $\Delta\ell =0$, $P_d = 200$ mW, and different coupling strength of the atomic and cavity fields: $\sum_jg(\rho_j)= 0.01\gamma_e$, $\sum_jg(\rho_j)= 0.05\gamma_e$, $\sum_jg(\rho_j)= 0.1\gamma_e$ and $\sum_jg(\rho_j) = 0.15\gamma_e$, respectively.
  Correspondingly, the case with weak $\Omega_{\beta}^{(\mathrm{max})}=0.02\times 2\pi$ MHz is shown in (b$_1$) and (b$_2$) with the same parameters.}
  \label{Fig_OMIT1}
\end{figure}

Figure \ref{Fig_nodriving}(c) displays the absorptive coefficient as a function of normalized optical detuning $(\delta_p-\omega_{\phi})/\omega_{\phi}$ for varying OAM values $l$ of the cavity mode.~Activation of the extra-cavity driving field ($E_d \neq 0$, \emph{Active case}) results in a sharp decrease in the absorptive curve at the resonance frequency $\delta_p=\omega_{\phi}$, creating a transparent window, a signature of the OMIT effect \cite{EIT-theory, EIT-experiments}, similar to vibrating optomechanical systems. With an increase in OAM values $l$ of the cavity mode, the OMIT window widens, and the optomechanical coupling constant $g_{\phi}=cl /L$ in the L-G cavity system correspondingly strengthens, resulting in the broadening of the OMIT window. In this case, the absorption coefficient reads $\upsilon_p\simeq\frac{2\kappa}{\kappa-\mathrm{i}\delta_p-\frac{G^2|a_s|^2}{\gamma_{\phi}-\mathrm{i}\delta_p}}$ with effective optomechanical interaction rate $Ga_s$. 

\begin{figure}[ptb]
  \centering
  \includegraphics[width=0.48\textwidth]{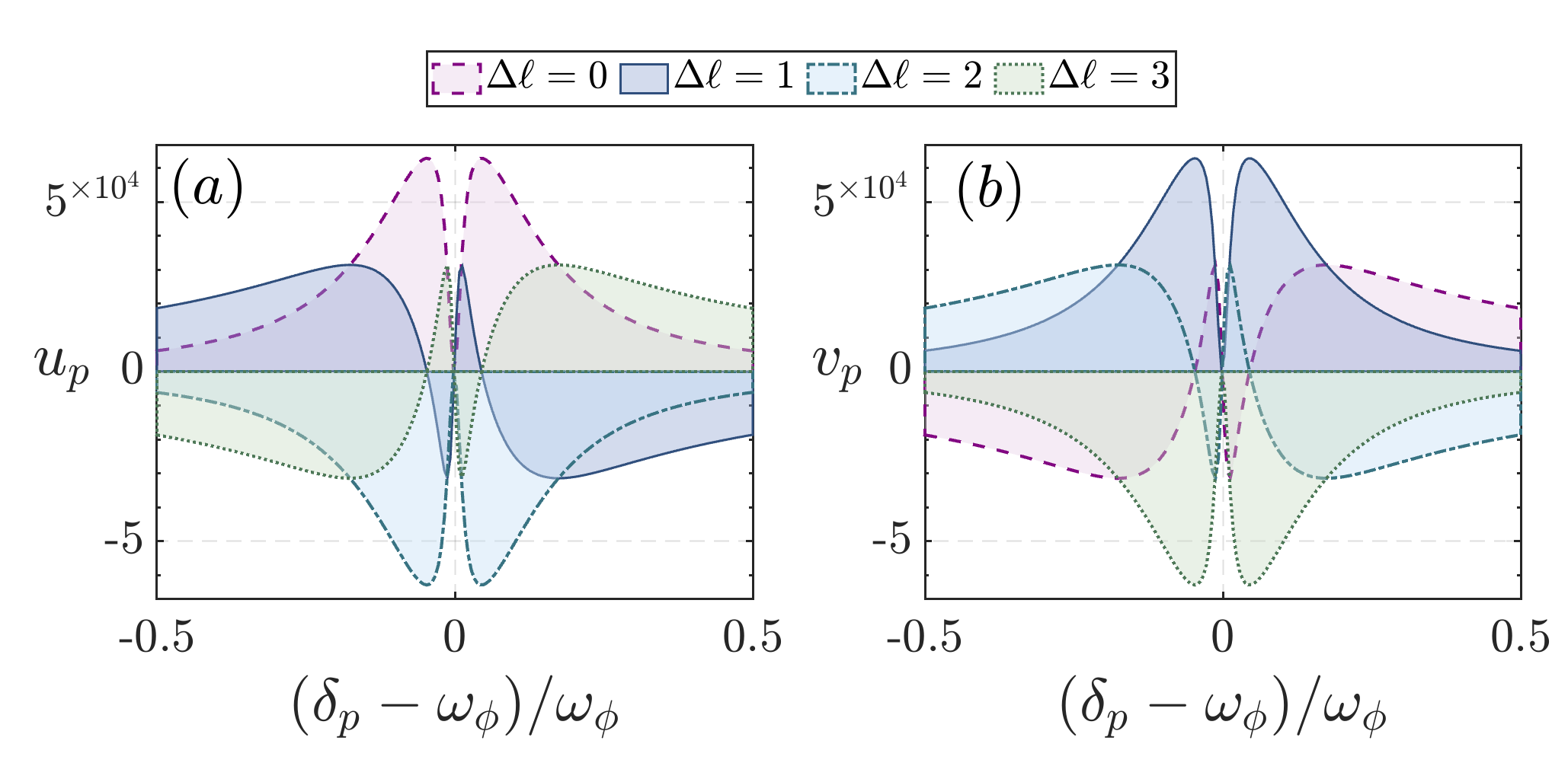}
  \caption{The absorption curves (a) and dispersion curves (b) as functions of the normalized optical detuning $(\delta_p-\omega_{\phi})/\omega_{\phi}$ with weak $\Omega_{\alpha}$ = 0.02$\times 2\pi$ MHz in the case of $\kappa=2.85\times2\pi$ MHz, $\sum_jg(\rho_j)=\gamma_e$, and different $\Delta\ell$: $\Delta\ell$ = 0, 1, 2 and 3. Other parameters are the same as Fig.~\ref{Fig_OMIT1}.}
\label{Fig_PT}
\end{figure}

The impact of atomic-cavity coupling strength on the absorption and dispersion characteristics of the output field with $\Delta\ell =0$ is depicted in Fig.~\ref{Fig_OMIT1}(a$_1$) and (a$_2$), under the condition of a weak field $\Omega_\alpha$. With increasing coupling strength $\sum_j g(\rho_j )$, the absorption window's peaks rise progressively, and the dispersion curve exhibits a pronounced steepness. In contrast to the $\Delta\ell=0$ scenario, when $\Delta\ell =2$, the system's output shows a trend of increasing optical gain ($v_p<0$) and enhanced normal dispersion ($\frac{\partial u_p}{\partial \omega}>0$) as the coupling strength augments, as illustrated in panels (b$_1$) and (b$_2$) of Fig.~\ref{Fig_OMIT1}. From Eq.~\eqref{Eq_delta_a4} and \eqref{Eq_out_T}, the output of the hybrid system reads 
\begin{align}
  \mathcal{E}_T=\frac{2\kappa[1-\mathrm{i}e^{\mathrm{i}\Delta\ell\theta}\sum_j\frac{g(\rho_j)\Omega_{\alpha}(\rho_j)\Omega_{\beta}^*(\rho_j) \Omega_{\zeta}(\rho_j)}{\gamma_e(|\Omega_\beta(\rho_j)|^2+|\Omega_\zeta(\rho_j)|^2)}]}{\kappa_{\mathrm{eff}}-\mathrm{i}\delta_p-\frac{G^2n_c}{\gamma_{\phi}-\mathrm{i\delta_p}}},
\end{align}
which predicts that different phase information ($\Delta\ell\cdot\theta$) leads to distinct absorption/gain ($v_p$) phenomena [\ref{B}] for $l=2m$. 

Figure~\ref{Fig_PT} demonstrates the possibility of engineering a unique symmetry relationship between the real and imaginary parts of the optical system in the frequency domain, under specific parameter settings \cite{PT-relationship-1,PT-relationship-2,PT-relationship-3, PT-relationship-4}. These symmetries can be modulated by the atomic system introduced phase, which satisfies $\Delta\ell \cdot\theta=n\pi, (n=0,1,2,...)$. Establishing a correspondence between the system's frequency domain and spatial domain, as in \cite{PT-1}, could enable the use of OAM within the system to achieve controllable parity-time symmetry ($\mathcal{PT}$) or parity-time antisymmetry ($\mathcal{APT}$) \cite{PT-2, PT-3, PT-4, PT-5}.

Furthermore, by altering the value of $\Delta \ell $, the transparency of the window and the slope of the dispersion curve will be adjustable, which also indicates that the system allows for fast and slow light conversion.~The phase of the relevant output field at frequency $\omega_p$ ($\omega$) can be written as 
\begin{align}\label{Eq_phi}
  \Theta(\omega)=\mathrm{arg}[t_p(\omega)]. 
\end{align}
Here, the rapidly varying dispersion in the narrow transparency window may cause the group delay that can be expressed as \cite{EIT-experiments, fast-slow-light-1},
\begin{align}\label{Eq_tau}
  \tau_g=\frac{\partial\Theta}{\partial\omega}. 
\end{align}

\begin{figure}[ptb]
  \centering
  \includegraphics[width=0.48\textwidth]{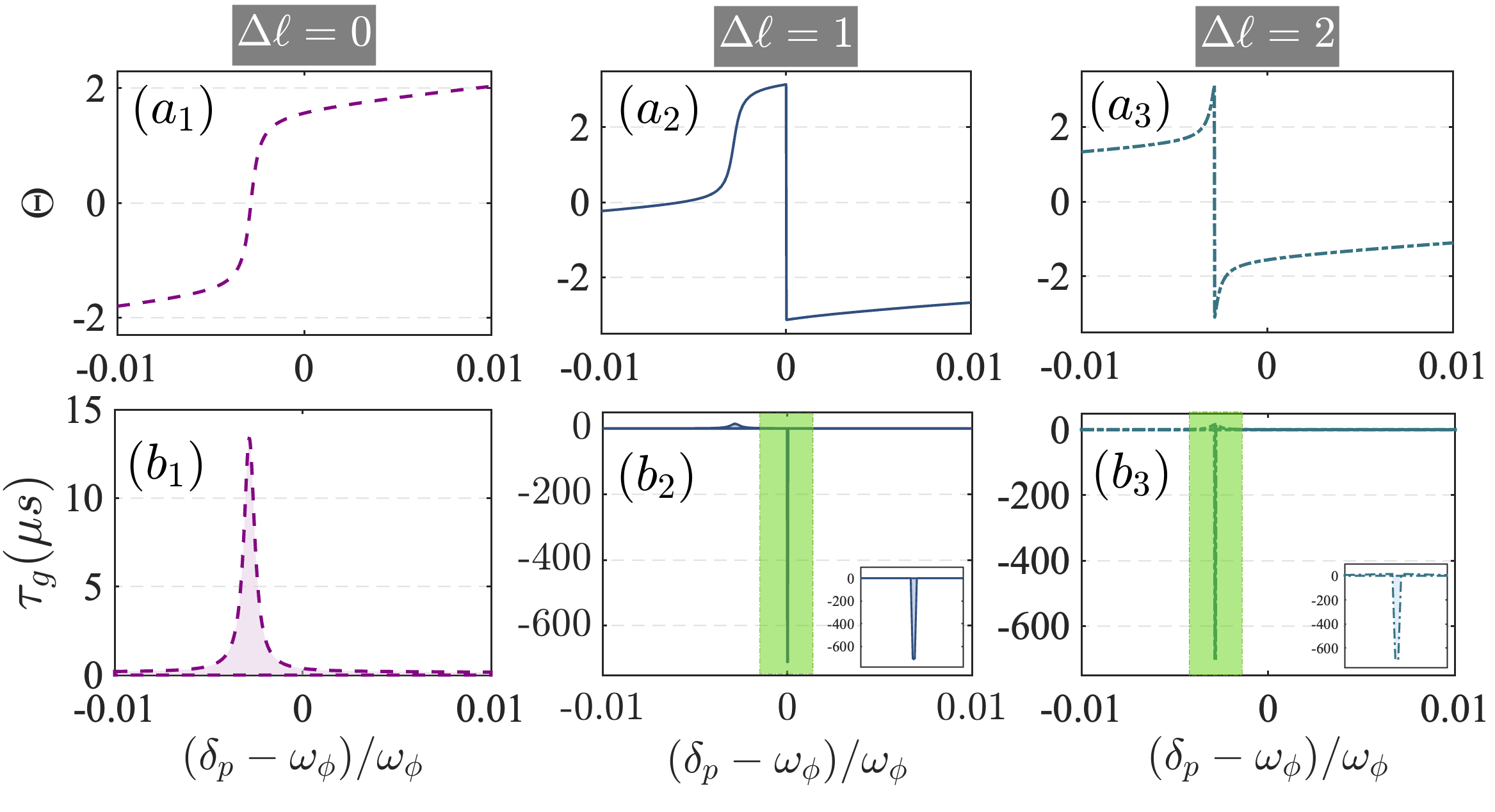}
  \caption{The output phase and group delay curves under different $\Delta \ell $, are shown in panels (a$_1$)-(a$_3$) and panels (b$_1$)-(b$_3$), respectively. The subpanels correspond to the green range of panels (b$_2$) and (b$_3$). The other parameters are the same as in Fig.~\ref{Fig_PT}. }
\label{Fig_delay1}
\end{figure}

As the support, curves for the output field phase (the group delay for the system) versus relative probe frequency $(\delta_p-\omega_{\phi})/\omega_{\phi}$ are displayed in Fig.~\ref{Fig_delay1} (a$_1$)-(a$_3$) [Fig.~\ref{Fig_delay1} (b$_1$)-(b$_3$)] with different $\Delta\ell$, near the resonant region. With null atomic OAM incidence ($\Delta\ell=0$), the system exhibits normal dispersion, characterized by a consistently positive slope in phase or dispersion [see Fig.~\ref{Fig_delay1}(a$_1$)].~Correspondingly, a time group delay peak around $(\delta_p-\omega_\phi)/\omega_\phi=-0.0028$ is observed, which corresponds to a slow light generation ($\tau_{g}=14.0$ $\mu$s), as shown in Fig.~\ref{Fig_delay1}(b$_1$).~However, with $\Delta\ell=1$ or 2, the phase dispersion changes rapidly (anomalous dispersion $\frac{\partial \Theta}{\partial \omega}<0$ appearing), which leads to large negative group velocity time delays (fast light appearing). It is noteworthy that the case with $\Delta\ell = 2$ generates fast light at the same frequency position ($\delta_p/\omega_\phi=0.9972$) as the case with $\Delta\ell = 0$. The system group delay up to $-698$ $\mu$s, which is a considerable improvement over the previous work with the L-G cavity \cite{L-G-group-decay-1}. The proposed method readily facilitates the transition between fast and slow light propagation by merely altering the difference in orbital angular momentum, $\Delta\ell\hbar$, among the three classical coherence fields. It is noted that our investigation of fast/slow light switching typically focuses on specific frequencies (such as the frequency $\delta_p/\omega_\phi=0.9972$ above), since maintaining fast/slow light characteristics across a broad bandwidth is inherently challenging. For instance, in Fig.~\ref{Fig_delay1}, slow light persists on both sides of frequency when $\Delta \ell$ = 1 or 2.

\begin{figure}[ptb]
  \centering
  \includegraphics[width=0.48\textwidth]{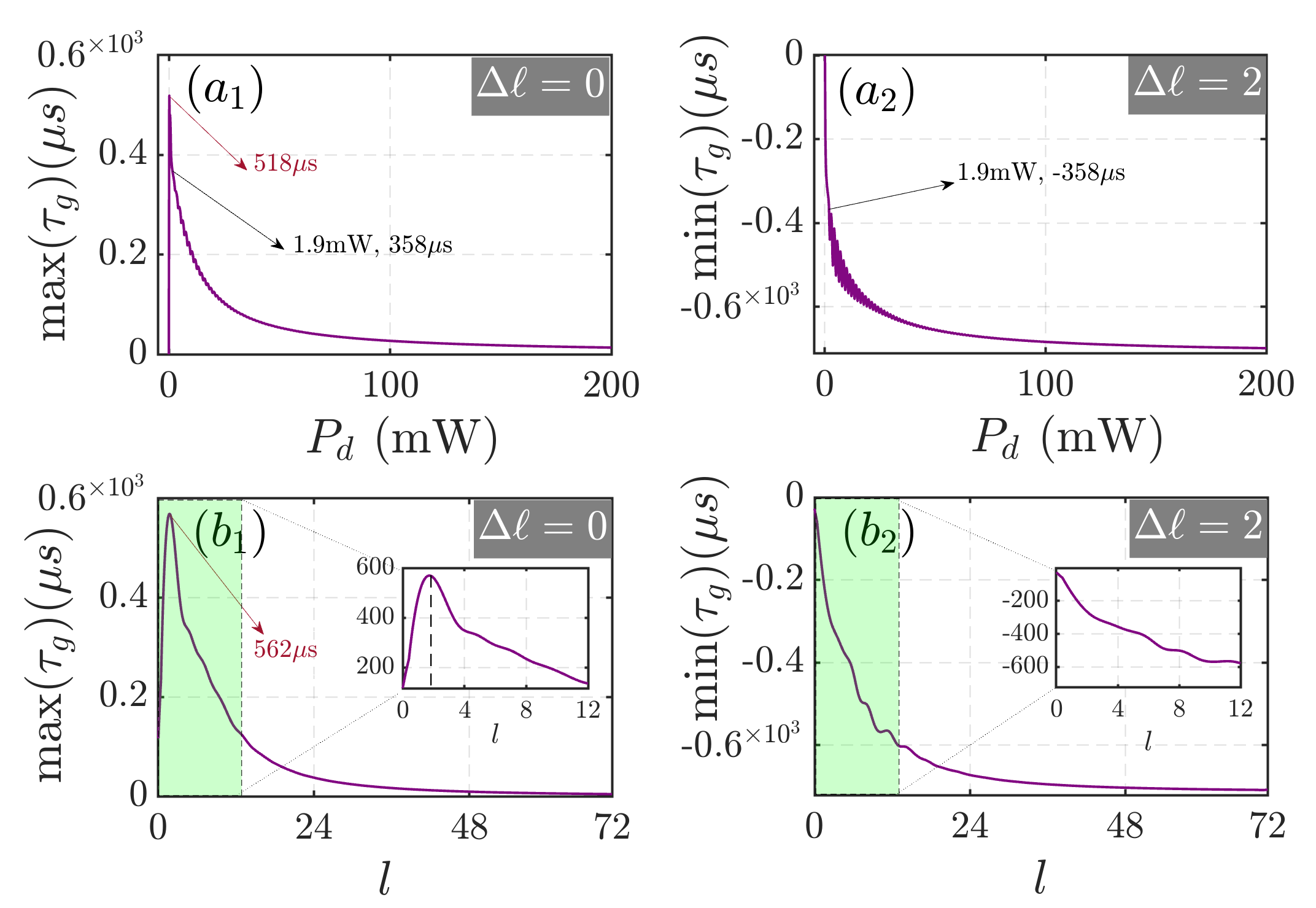}
  \caption{Maximum transmission group delay $\tau_g^{(\mathrm{max})}$ (a$_1$) and  minimum negative group delay $\tau_g^{(\mathrm{min})}$ (a$_2$) as functions of the driving field power $P_d$ for $\theta=\pi/2$. Correspondingly, the maximum transmission group delay $\tau_g^{(\mathrm{max})}$ (a$_1$) and minimum negative group delay $\tau_g^{(\mathrm{min})}$ (a$_2$) as functions of $l$ are shown in (b$_1$) and (b$_2$) with the same parameters. The rest of the parameters are the same as in Fig.~\ref{Fig_PT}.}
  \label{Fig_delay2}
\end{figure}

After the analysis of the effects of atomic nonlinearity on group delay, we proceed to investigate additional mechanisms for manipulating group delay. These include the optomechanical coupling constant, given by $g_{\phi} = \frac{cl}{L}$, and the driving intensity, expressed as $\mathcal{E}_d = \sqrt{\frac{2P_d \kappa}{\hbar\omega_d}}$
.~The maximum slow/fast light that can be produced by the system, under varying driving field powers [shown in Fig.~\ref{Fig_delay2} (a$_1$) and (a$_2$)] and different vortex topological charge $l$ induced by SPP [shown in Fig.~\ref{Fig_delay2} (b$_1$) and (b$_2$)].
In the scenarios with $l=40$, the maximum group delay reaches 518 $\mu$s as the driving field power approaches zero, and the group delay decreases as the driving power further increases.~What's more, the minimum negative group delay varies from 0 $\mu$s to $-$698 $\mu$s as $P_d$ varies within the range of $(0, 200)$ mW.~With driving power $P_d=$ 1.9 mW, a similar degree of reduction and increase in group velocity can be achieved by changing only $\Delta\ell$ ($\ell_{\Theta}$).
When the driving power is 200 mW, the modulation of $l$ on fast/slow light is similar to tuning the driving power $P_d$.~The delay reaches a maximum of 562 $\mu$s as $l=2$ (the insert of Fig. \ref{Fig_delay2} (b$_1$)), and the minimum negative group delay gradually increases to 707 $\mu$s with $l$ increasing from 0 to 70 as shown in Fig.~\ref{Fig_delay2} (b$_2$). Certainly, we recognize that as $l $ increases, the edge scattering effects of L-G modes become more pronounced, which poses significant experimental challenges.

\subsection{Optical vortices in the output field}
We recognize two fundamental observations regarding manipulating orbital angular momentum in our system.~Firstly, a vortex cavity system enhances the input Laguerre-Gaussian beam with an extra $2l\hbar$ of OAM at the output stage. Secondly, the interaction with an atomic ensemble capable of FWM further modulates the cavity mode's OAM by an increment of $\Delta\ell\hbar$ ($\ell_{\Theta}\hbar$). These considerations render a discussion on the optical vortices present in the output field both pertinent and significant.~The operator of the cavity field is redefined so that it carries phase information (OAM):
\begin{align}\label{Eq_c}
  a\rightarrow\tilde{a}=\tilde{a}_s+\delta\tilde{a}=a_s e^{2\mathrm{i}l\theta}+\delta a e^{2\mathrm{i}l\theta}. 
\end{align}
Accordingly, the form of the total output field can be expressed as
\begin{align}\label{Eq_in_out_re_piao}
  \tilde{\mathcal{E}}_{out}=\tilde{\mathcal{E}}_{out}^s+\tilde{\mathcal{E}}_{out}^{+}\mathcal{E}_{p}e^{-\mathrm{i}\delta_p t}+\tilde{\mathcal{E}}_{out}^{-}\mathcal{E}_{p}e^{\mathrm{i}\delta_p t},
\end{align}
where 
\begin{align}\label{Eq_out_piao}
  \tilde{\mathcal{E}}_{out}^s= & 2\kappa a_{s} e^{2\mathrm{i}l\theta}-\mathcal{E}_{d}, \nonumber\\
  \tilde{\mathcal{E}}_{out}^{+}= & 2\kappa \delta a_{+} e^{2\mathrm{i}l\theta}-1, \nonumber\\
  \tilde{\mathcal{E}}_{out}^{-}= & 2\kappa \delta a_{-} e^{2\mathrm{i}l\theta}. 
\end{align}

\begin{figure}[ptb]
  \centering
  \includegraphics[width=0.48\textwidth]{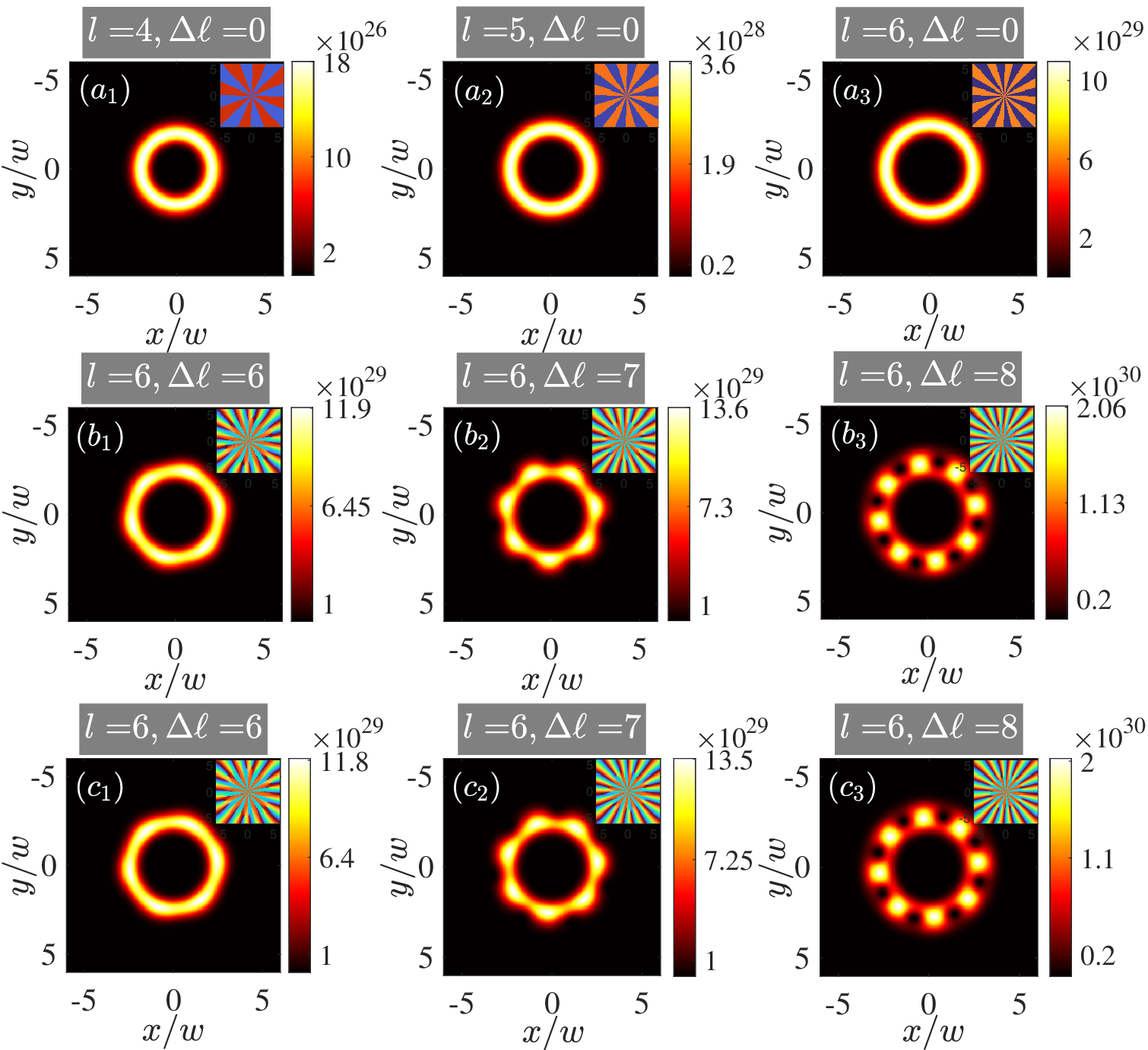}
  \caption{The spot patterns of the total output field: Absence of OAM from the atom part ($\Delta\ell\hbar=0$), the L-G cavity carries different OAM $l\hbar$ ($l=$ 4, 5 and 6) in panel (a$_1$), (a$_2$) and (a$_3$), respectively. 
  The coupling fields carry different OAM (b$_1$), (b$_2$) and (b$_3$) with $\delta_p=\delta_c=0$; Panel (c$_1$), (c$_2$) and (c$_3$) display the cases with $\delta_{c}=4\times 2\pi$MHz. The inside panels show the spatial distribution of the phases. The coupling strength between the atom and cavity field is set as $\sum_jg(\rho_j)=10^{-2}\times 2\pi$ MHz, with other parameters being the same as in Fig.~\ref{Fig_OMIT1}.}
  \label{Fig_OAM1}\centering
\end{figure}  

We first consider the situation with a nonzero detuning between the driving field and the cavity mode ($\Delta_c\neq 0$). In this scenario, for the probe frequency, the total output spot's complex amplitude obtained on the left side of the hybrid system should be denoted as $\mathcal{E}=\eta\mathcal{E}_p e^{-2\mathrm{i}l\theta}+\tilde{\mathcal{E}}_{out}$. Here, the first term represents the portion of the probe light reflected by the left cavity wall carrying an OAM of $-2l\hbar$ after scattering by SPP, with whose reflective rate $\eta$. The other part consists of the output cavity mode of the hybrid system.~With a weak coupling strength between the atom and the cavity field ($\sum_jg(\rho_j)=10^{-2}\times\gamma_e$), the pattern of the output spot of the total output field is shown in Fig.~\ref{Fig_OAM1}. The diameter of the Laguerre-Gaussian beams' ring pattern increases with the augmentation of $l$, with $\Delta\ell$ set to 0, as depicted in Fig.~\ref{Fig_OAM1} (a$_1$), (a$_2$), and (a$_3$). Given the extremely weak probe intensity, the driving field and the atomic component dominate the final output beam profile. Upon incorporating the atomic manipulation, the L-G beams' intensity pattern exhibits periodic interference rings, as illustrated in Fig.~\ref{Fig_OAM1}(b$_1$), (b$_2$), and (b$_3$), with the number of periods equal to $\Delta\ell$, while keeping $l$ constant. The three bottom panels demonstrate the situation where the atom is detuned from the cavity mode $(\delta_c\neq 0)$. Apart from a slight decrease in the intensity of the beams, it is almost identical to the resonant case $(\delta_c = 0)$. Certainly, the phase diagrams presented in the sub-figures, ranging from Fig.~\ref{Fig_OAM1} (a$_1^{\mathrm{in}}$) to Fig.~\ref{Fig_OAM1} (c$_3^{\mathrm{in}}$), provide strong support for our conclusions.

\begin{figure}[ptb]
\centering
\includegraphics[width=0.48\textwidth]{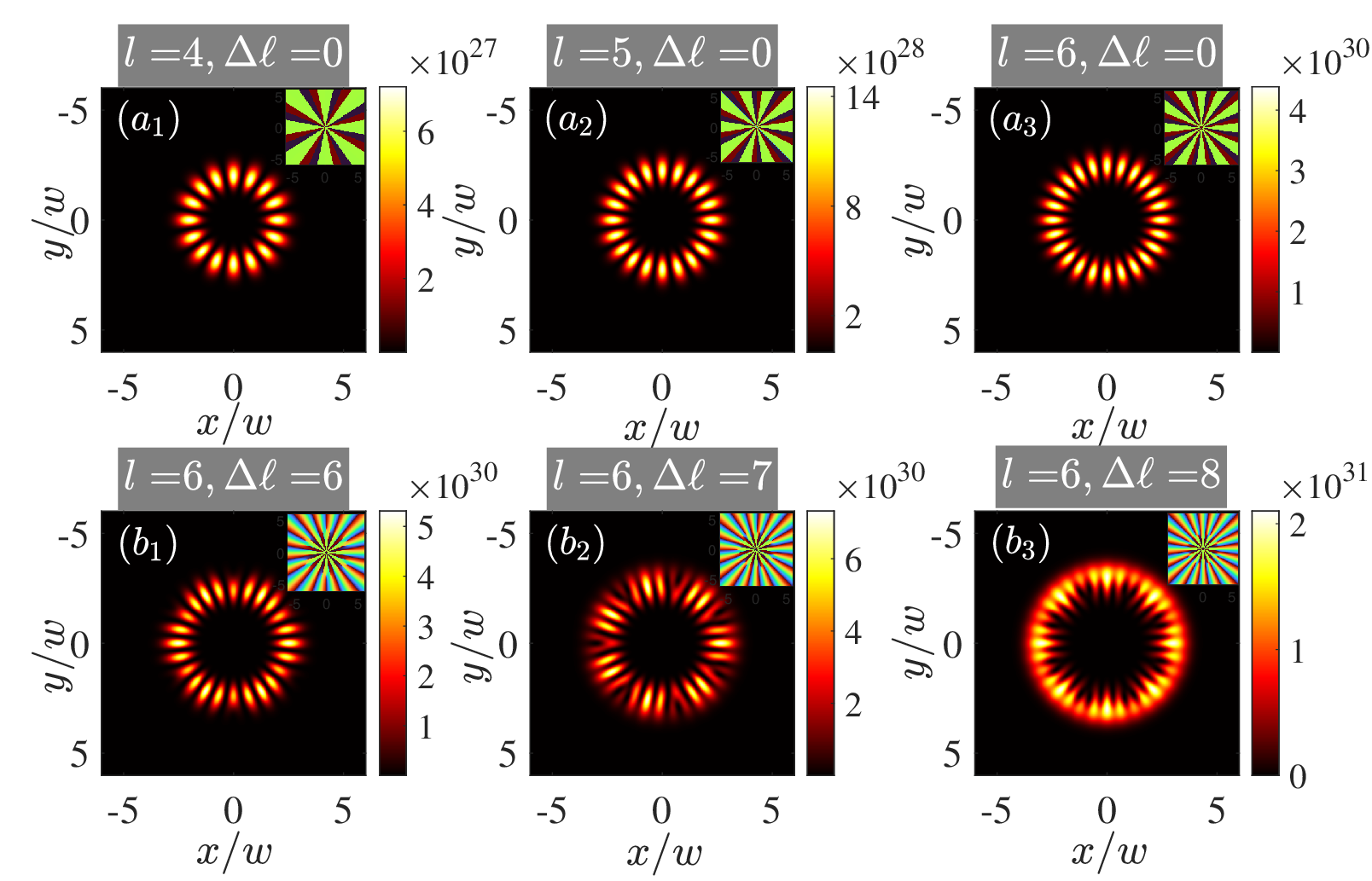}
  \caption{The spot patterns of the total output field $\tilde{\mathcal{E}}_{out}^s$ as a function of $(x,y)$, with $\omega_p=\omega_d=\omega_c$. The other parameters used here are the same as in Fig.~\ref{Fig_OAM1}.}
  \label{Fig_OAM2}\centering
\end{figure}  

In another scenario, when the driving field, probe field, and cavity mode are all resonant $(\Delta_c = \delta_p = 0)$, the total output field should be the superposition of three components, $\mathcal{E}=\eta\mathcal{E}_p e^{-2\mathrm{i}l\theta}+\eta\mathcal{E}_d e^{-2\mathrm{i}l\theta}+\tilde{\mathcal{E}}_{out}$. Here, compared to the first scenario, there is an additional component due to the reflection of the driving field by the left SPP, denoted as $\eta\mathcal{E}_d e^{-2\mathrm{i}l\theta}$, which carries an OAM of $-2l\hbar$. The total output light intensity is shown in Fig.~\ref{Fig_OAM2}(a$_1$), (a$_2$), and (a$_3$), whose patterns are petal-like, and the period of the petals is $2\times 2|l| $. The atoms contribute little to the total number of outgoing fields, so that the total output field can be approximated as a superposition of two vortex beams of different intensities with OAMs of $ +2l\hbar$ and $-2l\hbar$, respectively. However, with increasing the coupling strength between the atoms and the cavity field to $\sum_j g(\rho_j )=10^{-2}\times\gamma_e$, the vortex phase factor $ \ell_{\Theta}$ changes the pattern of the total output light intensity, as shown in Fig.~\ref{Fig_OAM2}(b$_1$), (b$_2$) and (b$_3$). The number of petals is still $2\times 2|l| $, but the number of pedal periods is regulated by the atomic part $\Delta\ell$ ($ \ell_{\Theta}=\Delta\ell+2l$).

\subsection{Effects of mode mismatch and noise}
In the above analysis, we assumed that the cavity mode and the driving field are perfectly aligned, which is not always the case. The mismatch between the cavity mode and the driving field can affect the system's output characteristics and result in a decrease in effective phase control. In this section, we will discuss how mode mismatch affects the output characteristics of the hybrid system.

\begin{figure}[ptb]
  \centering
  \includegraphics[width=0.48\textwidth]{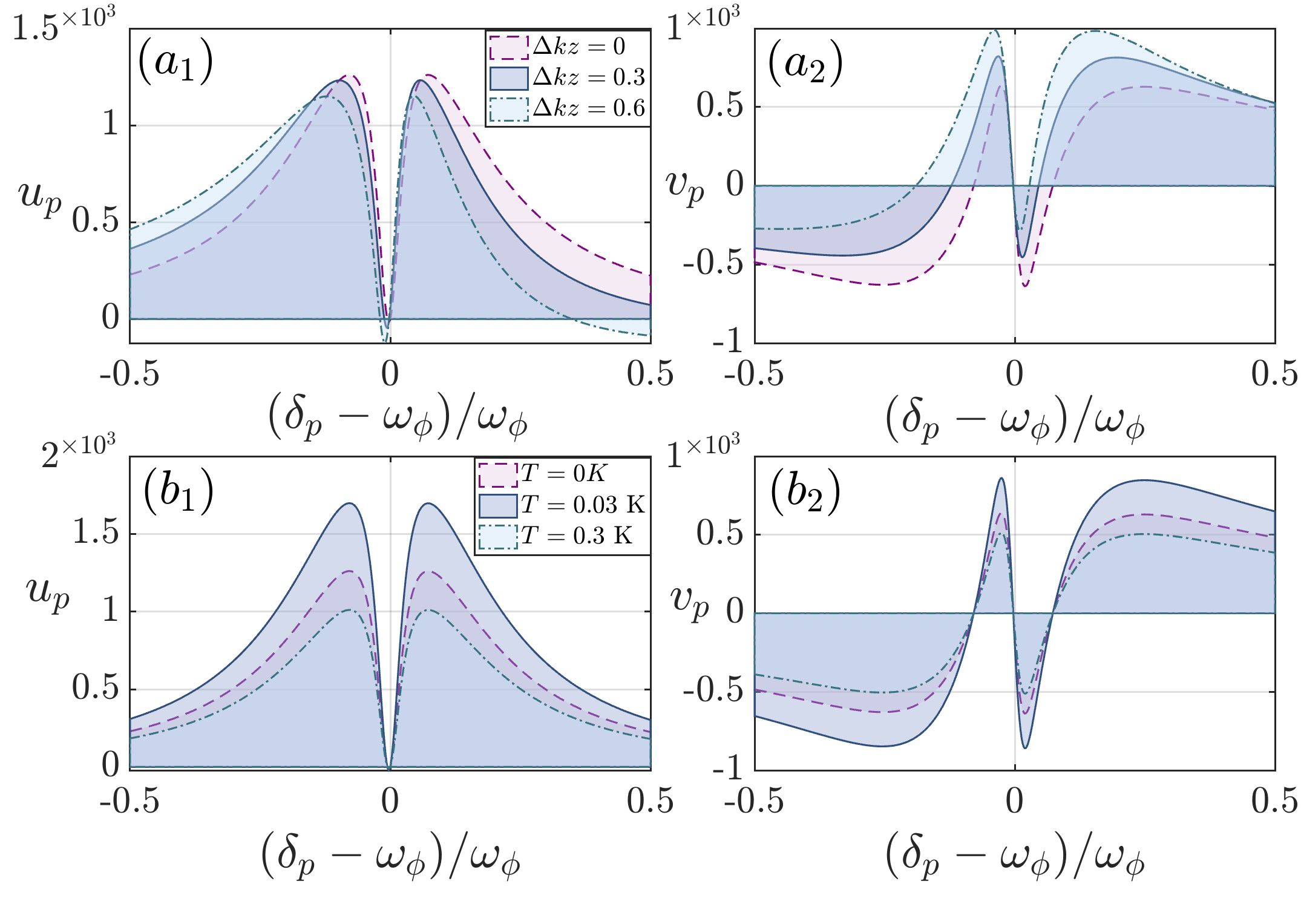}
  \caption{The total output's absorption and dispersion curves versus frequency under two types of modifications: (a$_1$) and (a$_2$) correspond to absorption and dispersion under different phase mismatch conditions, while (b$_1$) and (b$_2$) represent absorption and dispersion with different temperature corrections. All other parameters remain identical to those in Fig.~\ref{Fig_OMIT1}, expect for $\Delta \ell=0$, $\sum_jg(\rho_j)=0.01\gamma_e$.}
  \label{fig10}\centering
\end{figure}

First of all, the dominant effect in the atomic FWM process arises from phase mismatch, defined as $\Delta k = \overrightarrow{k_{\alpha}} - \overrightarrow{k_{\beta}} + \overrightarrow{k_{\zeta}} - \overrightarrow{k_{c}}$, which is incorporated into the modified coefficient $B_0 \to B_0^{mm} = \frac{\mathrm{i} \Omega_\alpha \Omega_\beta^* \Omega_\zeta e^{\mathrm{i}(\Delta k \cdot z - \Delta\ell \theta)}}{\delta_{c}^2 \gamma_e - (|\Omega_\beta|^2 + |\Omega_\zeta|^2) \gamma_e - \mathrm{i} \delta_c (|\Omega_\beta|^2 + \gamma_e^2)}$. Here, the weak-field case with $\Omega_{\alpha}$ exemplifies this phase-mismatch mechanism. Figure \ref{fig10}($a_1$) and ($a_2$) depict the output absorption and dispersion spectra of the hybrid system for different values of $\Delta k\cdot z$. This phase mismatch reduces the effective phase control, manifesting as a broadening of the absorption spectrum. Given the system size with maximum length $z_{\text{max}} = L$, we estimate that the tolerance for phase mismatch should be maintained below $$|\Delta k / k_c| = 10^{-3}$$ to ensure the effectiveness of $\Delta\ell$ in slow and fast light modulation. The basis for estimating this limit is the condition $\Delta k \cdot L \leq \pi/2$, corresponding to $\phi = \Delta k \cdot z_{\text{max}} + \ell_{\Theta} \pi/2$ with $\Delta\ell = 0$. Under these conditions, phase mismatch induces a significant deviation in phase modulation.

Next, spatial mode matching between the cavity mode and the classical beams, which interact with atoms via carefully aligned L-G beams, is achieved by maintaining nearly identical spatial distribution parameters across all beams. Specifically, the beam waist radius satisfies $w_{c0} \simeq w_\mu = w_0$ with $\mu \in \{\alpha, \beta, \zeta\}$, while the spatial indices $l$ and $p$ remain consistent. Crucially, all classical beams ($\alpha$, $\beta$, and $\zeta$) propagate at small angles $\phi_\mu$ nearly parallel to the optical axis. This angular alignment results in an effective beam waist perpendicular to the optical axis, given by $w_\mu^\perp = w_0 / \cos\phi_\mu$. Due to the small magnitude of these angles, $\cos\phi_\mu \approx 1$, yielding the approximation $w_\mu^\perp \simeq w_0$. Collectively, these spatial and angular conditions minimize errors from the spatial mode mismatch in the system.

Moreover, the noise in the system can also affect the output characteristics. The noise can be introduced by the thermal fluctuations of the cavity mode and the Doppler effect of the atomic ensemble, which can lead to a reduction in the effective phase control. The standard computational approach involves performing temperature-dependent integration over the atomic component, $\varrho_{14}^{D}=\int\rho_{14}(v)M(v)dv$ with $M(v)=(\mu\sqrt{\pi})^{-1}\exp(-v^2/\mu^2)$ and $\mu=\sqrt{2k_BT/m}$. Since our calculation requires obtaining an analytical solution for the atomic part, we employ a Lorentzian distribution ($L=\frac{1}{\mu\sqrt{\pi}+\frac{2\sqrt{\pi}v^2}{\ln{2}}}$) approximating the Maxwellian distribution to account for finite-temperature corrections \cite{Doppler}.
Therefore, the temperature-corrected atomic component can be expressed as,
\begin{widetext}
\begin{align}
 A_0' &=\frac{g(\rho)\sqrt{\pi\ln{2}}[(\delta_c+\mathrm{i}k\sqrt{\frac{\mu\ln{2}}{2}})\gamma_e-\mathrm{i}|\Omega_\beta(\rho)|^2]}{\sqrt{2\mu}(\delta_c+\mathrm{i}k\sqrt{\frac{\mu\ln{2}}{2}})^2\gamma_e-\sqrt{2\mu}(|\Omega_\beta(\rho)|^2+|\Omega_\zeta(\rho)|^2)\gamma_e-\mathrm{i}\sqrt{2\mu}(\delta_c+\mathrm{i}k\sqrt{\frac{\mu\ln{2}}{2}})(|\Omega_\beta(\rho)|^2+\gamma_e^2)},\nonumber\\
 B_0'&=\frac{\mathrm{i}\sqrt{\pi\ln{2}}\Omega_\alpha(\rho) {\Omega_\beta}^*(\rho) \Omega_\zeta(\rho)e^{\mathrm{i}\Delta\ell\pi/2}}{\sqrt{2\mu}(\delta_c+\mathrm{i}k\sqrt{\frac{\mu\ln{2}}{2}})^2\gamma_e-\sqrt{2\mu}(|\Omega_\beta(\rho)|^2+|\Omega_\zeta(\rho)|^2)\gamma_e-\mathrm{i}\sqrt{2\mu}(\delta_c+\mathrm{i}k\sqrt{\frac{\mu\ln{2}}{2}})(|\Omega_\beta(\rho)|^2+\gamma_e^2)}.
\end{align}
\end{widetext}

Figure \ref{fig10} demonstrates the influence of the Doppler effect on the OMIT (optomechanically induced transparency) lineshape. Therefore, finite temperature only affects the linewidth and dispersion values of OMIT (optomechanically induced transparency) without altering its overall trend, thus leaving the qualitative conclusions regarding $\Delta\ell$-dependent phase control in the preceding analysis unaffected.

Additionally, mechanical noise primarily originates from two sources: thermal noise ($\sigma_{\psi}$ induced by ambient temperature $T$) and quantum noise (zero-point fluctuations $\psi_{\text{zpf}}$, which dominate at low temperatures). For a rotational oscillator, the power spectral density (PSD) of angular displacement noise is dominated by thermal fluctuations. This takes a form analogous to that of a translational oscillator 	\cite{PSD} but with the moment of inertia $I$ substituted for the mass $m$:  
\begin{equation}  
S_{\phi\phi}(\omega) = \frac{4k_B T \gamma_{\phi}}{I[(\omega_m^2 - \omega^2)^2 + \gamma_{\phi}^2 \omega^2]}.  
\end{equation}  
When an optical field carries orbital angular momentum (OAM), the relationship between phase deviation $\delta\psi$ and angular displacement $\delta\phi$ is given by $\delta\psi = l\delta\phi$. Consequently, $S_{\psi\psi}(\omega) = l^2 S_{\phi\phi}(\omega)$. The root mean square (rms) phase error is expressed as  
\begin{equation}  
\sigma_{\psi} = \sqrt{\int_{-\infty}^{\infty} S_{\psi\psi}(\omega)  d\omega} = |l| \sqrt{\frac{k_B T Q_m}{I \omega_m^3}},  
\end{equation}  
which depends primarily on $|l|$ and $T$. However, quantum noise (zero-point fluctuations) dominates at low temperatures (e.g., $T < 100$ mK), as characterized by  $\psi_{\text{zpf}} = \sqrt{\frac{\hbar}{2I\omega_m}}$.

\section{Conclusion}\label{SecIV}
In summary, we have investigated the steady-state optical response of a hybrid system composed of a Laguerre-Gaussian vortex cavity and ultracold atoms driven into a double-$\Lambda$ model. The atomic ensemble, due to the FWM process involving three vortex beams carrying OAMs ($\hbar\ell_{\alpha,\beta,\zeta}$), is induced to have a loop phase $\varphi =\ell_{\Theta}\theta$ ($\ell_{\Theta}=\Delta\ell+2l$ and $\Delta\ell=\ell_{\alpha}-\ell_{\beta}+\ell_{\zeta}$). Initially, we analyzed the impact of two distinct four-wave mixing (FWM) coupling schemes on the OMIT spectrum of the hybrid system. It is discovered that the net OAM ($\ell_{\Theta} \hbar$) of the closed loops formed by the atomic FWM system can switch the system's output characteristics between absorption ($\Delta\ell=0$) and gain ($\Delta\ell=2$), with $\theta=\pi/2$ and $l=2m$ (here $ e^{\mathrm{i}\Delta\ell\pi/2}=e^{\mathrm{i}\ell_{\Theta}\pi/2}$). Correspondingly, the group delay characteristics (fast/slow light) of the output beam from the hybrid system are also controlled and switched by $\Delta\ell\hbar$. Subsequently, we found and discussed the periodicity of the output spot under various detuning conditions between the driving frequency, cavity mode frequency, and probe frequency. Meanwhile, phase control is sensitive to errors. To suppress errors induced by the Doppler effect and mechanical noise, a smaller $l$ ($|l|<10$) and lower temperature are necessary ($T< 100$ mK). Additionally, the control of the OAM and phase mismatch of the L-G beam needs to be maintained  at relatively low levels ($|\Delta k/k_c|\ll10^{-3}$).

This work provides insights into the modulation of L-G optomechanical cavities and the application of OAM in hybrid quantum optical systems. Based on this composite system, it is also possible to discuss the transfer and control of OAM in the atomic system by the L-G mode in return, in which case the L-G cavity can be regarded as a special kind of OAM source. In the passive case (without cavity driving), combining vacuum fluctuations or multiple baths with different temperatures not only makes the situation more interesting but also gives rise to novel phenomena worth exploring, such as quantum heat engines with harmonic rotation.

\section*{Acknowledgment}
This work is supported by the Scientific Research Project of Jilin Provincial Department of Education (JJKH20241411KJ);  Jilin Scientific and Technological Development Program (20220101009JC); National Natural Science Foundation of China (12104107).

\appendix
\section{Atomic four-wave mixing and steady-state}\label{A1}
After the frame rotation in Eqs.~\eqref{Eq_rotating}, the interaction Hamiltonian for single atom reads as 

\begin{align}\label{H_af_rotating}
\tilde{\mathcal{H}}_{af} & =  -\hbar\begin{bmatrix} 
0               & 0 & \Omega_{\alpha}  & g_0 \hat{a}           \\
0        & \delta_{42}-\delta_c       & \Omega_{\beta}^{*}   & \Omega_{\zeta}              \\
\Omega_{\alpha}^{*}              & \Omega_{\beta}        & -\delta_{31}   & 0      \\
g_0\hat{a} ^{\dag}  & \Omega_{\zeta}^{*}  & 0          & -\delta_{c}   
\end{bmatrix},
\end{align}

Moreover, the Liouville equation is given by
\begin{align}\label{Eq_liouville_af}
\dot{\rho}_{12} & = -\gamma_{12}^{\prime}\rho_{12}-\mathrm{i}\Omega_{\beta}^{*}\rho_{13}-\mathrm{i}\Omega_{\zeta}\rho_{14}+\mathrm{i}g_0\hat{a}\rho_{42}, \nonumber\\
\dot{\rho}_{13} & = -\gamma_{13}^{\prime}\rho_{13}-\mathrm{i}\Omega_{\beta}^{*}\rho_{12}+\mathrm{i}\Omega_{\alpha}(\rho_{33}-\rho_{11}) \nonumber\\
& +\mathrm{i}g_0\hat{a}\rho_{43}, \\
\dot{\rho}_{14} & = -\gamma_{14}^{\prime}\rho_{14}-\mathrm{i}\Omega_{\zeta}\rho_{12}-\mathrm{i}\Omega_{\alpha}\rho_{34}+\mathrm{i}g_0\hat{a}(\rho_{44}-\rho_{11}). \nonumber
\end{align}

So we can obtain the steady-state of the atomic four-wave mixing system by solving the equation
$\rho_{14}^s=\frac{\mathrm{i}g_0\hat{a}(\gamma_{12}^{\prime}\gamma_{13}^{\prime})(\rho_{44}-\rho_{11})}{\gamma_{12}^{\prime}\gamma_{13}^{\prime}\gamma_{14}^{\prime}+\gamma_{14}^{\prime}|\Omega_{\beta}|^2+\gamma_{13}^{\prime}|\Omega_{\zeta}|^2}-\frac{\mathrm{i}\Omega_{\alpha}\Omega_{\beta}^{*}\Omega_{\zeta}(\rho_{33}-\rho_{11})}{\gamma_{12}^{\prime}\gamma_{13}^{\prime}\gamma_{14}^{\prime}+\gamma_{14}^{\prime}|\Omega_{\beta}|^2+\gamma_{13}^{\prime}|\Omega_{\zeta}|^2}$, if $\Omega_{\alpha}\ll\Omega_{\beta}\sim\Omega_{\zeta}$, with effective decay rates $\gamma_{12}^{\prime}=\gamma_{12}+\mathrm{i}(\delta_c-\delta_{42})$, $\gamma_{13}^{\prime}=\gamma_{13}+\mathrm{i}\delta_{31}$, and $\gamma_{14}^{\prime}=\gamma_{14}+\mathrm{i}\delta_{c}$.~Here,  we can divide $\rho_{14}^s$ into the linear part $\rho_{14}^{lin}= \frac{\mathrm{i}g_0\hat{a}(\gamma_{12}^{\prime}\gamma_{13}^{\prime})(\rho_{44}-\rho_{11})}{\gamma_{12}^{\prime}\gamma_{13}^{\prime}\gamma_{14}^{\prime}+\gamma_{14}^{\prime}|\Omega_{\beta}|^2+\gamma_{13}^{\prime}|\Omega_{\zeta}|^2}\simeq A_0$ and the nonlinear part (FWM) $\rho_{14}^{nl}=\frac{\mathrm{i}\Omega_{\alpha}\Omega_{\beta}^{*}\Omega_{\zeta}(\rho_{33}-\rho_{11})}{\gamma_{12}^{\prime}\gamma_{13}^{\prime}\gamma_{14}^{\prime}+\gamma_{14}^{\prime}|\Omega_{\beta}|^2+\gamma_{13}^{\prime}|\Omega_{\zeta}|^2}\simeq B_0$, if $\gamma_{12}^{\prime}\to 0$. We selected $^{87}$Rb atoms, whose spontaneous emission rates are $\Gamma_{41} \simeq \Gamma_{42} \simeq \Gamma_{31} \simeq \Gamma_{32} \simeq \Gamma_{\text{free}} \approx 6$ MHz. After accounting for the Purcell effect, the cavity-enhanced decay rate is given by
$\Gamma_c \simeq \Gamma_{\text{free}} \cdot (\frac{\lambda_0^3}{4\pi^2 V})$. Therefore, the coherent damping rates form as $\gamma_{mn}=\frac{\sum_m\Gamma_{mk}+\sum_n\Gamma_{nk}}{2}$ with $m,n \in\{1,2,3,4\}$ and $k<m,n$.

\section{Quantised cavity output field}\label{B}
Here, we analyze the steady-state output of the hybrid system from an analytical perspective. Taking an example with an obvious atomic nonlinearity ($B_{0}$), where $|\Omega_{\alpha}|\ll|\Omega_{\beta}|\sim|\Omega_{\zeta}|$ and $\delta_{c}= 0$, the correlation terms of the atomic suspectibility can be rewritten as
\begin{align}
  A_0 & =\frac{\mathrm{i}\sum_jg(\rho_j)|\Omega_{\beta}(\rho_j)|^2e^{\mathrm{i}2l\theta}}{\gamma_e[|\Omega_{\beta}(\rho_j)|^2+|\Omega_{\zeta}(\rho_j)|^2]} , \nonumber\\
  B_0 & =\frac{-\mathrm{i}e^{\mathrm{i}\Delta\ell\theta}\Omega_{\alpha}(\rho_j) \Omega_{\beta}^*(\rho_j) \Omega_{\zeta}(\rho_j)}{\gamma_e[|\Omega_\beta(\rho_j)|^2+|\Omega_\zeta(\rho_j)|^2]}.
\end{align}

According to Eq. \eqref{Eq_delta_a2}, the form of the output field becomes
\begin{align}\label{Eq_1}
  \mathcal{E}_T=\frac{2\kappa[1-\mathrm{i}\sum_j\frac{e^{\mp\mathrm{i}\Delta\ell\theta}g(\rho_j)\Omega_{\alpha}(\rho_j)\Omega_{\beta}^*(\rho_j) \Omega_{\zeta}(\rho_j)}{\gamma_e[|\Omega_\beta(\rho_j)|^2+|\Omega_\zeta(\rho_j)|^2]}]}{\kappa_{\mathrm{eff}}-\mathrm{i}\delta_p-\frac{G^2n_c}{\gamma_{\phi}-\mathrm{i\delta_p}}},
\end{align}
for $l=2m-1$ or $l=2m$ ($m\in\mathbb{Z} $), with the effective cavity field damping coefficient
\begin{align}\label{A3}
   \kappa_{\mathrm{eff}}=\kappa-\sum_j\frac{g(\rho_j)|\Omega_\beta(\rho_j)|^2}{\gamma_e[|\Omega_\beta(\rho_j)|^2+|\Omega_\zeta(\rho_j)|^2]}.
\end{align}

For summations involving spatially distributed terms, it is convenient to employ approximations to compute the average value.

If the output is split into $\mathcal{E}_T=u_p+\mathrm{i}v_p$ by the real and imaginary parts, then a discussion is needed on the values of $\Delta\ell\theta$. Three distinct cases are examined for $\Delta \ell=0,1,2$, under the assumption that $\theta = \pi/2$ and $l=2m$.~Taking $\Delta\ell=0$, firstly, we can easily get $e^{\mathrm{i}\Delta\ell\theta}=1$. According to Eq. \eqref{Eq_1}, the real and imaginary parts of the output field can be obtained separately as
\begin{widetext}
\begin{align}\label{Eq_0}  
  v_p^{\Delta\ell=0} & =\sum_j\frac{2\kappa\gamma_{e}\delta_p\mathcal{M}\mathit{\Omega}^2(\rho_j)[\gamma_e\mathit{\Omega}^2(\rho_j)-g(\rho_j)\Omega_{\alpha}(\rho_j)\Omega_{\beta}^{*}(\rho_j)\Omega_{\zeta}(\rho_j)]}{DD(\rho_j)}, \\  u_p^{\Delta\ell=0} & =\frac{2\kappa(\gamma_e\mathit{\Omega}^2(\rho_j)\mathcal{N}-g^{2}(\rho_j)\Omega_{\beta}^2(\rho_j)\mathcal{G})[\gamma_e\mathit{\Omega}^2(\rho_j)-g(\rho_j)\Omega_{\alpha}(\rho_j)\Omega_{\beta}^{*}(\rho_j)\Omega_{\zeta}(\rho_j)]}{DD(\rho_j)},\nonumber
\end{align}
\end{widetext}
where $\mathit{\Omega}^2=\Omega_{\beta}^2+\Omega_{\zeta}^2$, $\mathcal{G}=\gamma_{\phi}^2+\delta_p^2$, $\mathcal{M}=G^2n_c+\delta_p^2+\gamma_{\phi}^2$, $\mathcal{N}=\kappa\delta_p^2-G^2n_c\gamma_{\phi}+\kappa\gamma_{\phi}^2$, and $DD(\rho_j)=\sum_j \gamma_e^2\mathit{\Omega}^4(\rho_j)[\kappa^2\mathcal{G}+(\delta_p^2+G^2n_c)^2+\gamma_\phi^2\delta_p^2-2\gamma_\phi \kappa G^2n_c]-g^2(\rho_j )\Omega_{\beta}^2(\rho_j)[2\gamma_e\mathcal{N}\mathit{\Omega}^2(\rho_j)-g^2(\rho_j)\Omega_{\beta}^2(\rho_j)\mathcal{G}]$. 

Secondly, with $\Delta\ell=2$ so as to $e^{\mathrm{i}\Delta\ell\theta}=-1$, thus the real/imaginary parts of the output field will read as
\begin{widetext}
\begin{align}\label{Eq_2}
 v_p^{\Delta\ell=2} & =\sum_j\frac{2\kappa\gamma_{e}\delta_p\mathcal{M}\mathit{\Omega}^2(\rho_j)[\gamma_e\mathit{\Omega}^2(\rho_j)+g(\rho_j)\Omega_{\alpha}(\rho_j)\Omega_{\beta}^{*}(\rho_j)\Omega_{\zeta}(\rho_j)]}{DD(\rho_j)},\\  u_p^{\Delta\ell=2} & =\frac{2\kappa(\gamma_e\mathit{\Omega}^2(\rho_j)\mathcal{N}-g^{2}(\rho_j)\Omega_{\beta}^2(\rho_j)\mathcal{G})[\gamma_e\mathit{\Omega}^2(\rho_j)+g(\rho_j)\Omega_{\alpha}(\rho_j)\Omega_{\beta}^{*}(\rho_j)\Omega_{\zeta}(\rho_j)]}{DD(\rho_j)}.\nonumber
\end{align}
\end{widetext}
Similar to the above two cases, taking $\Delta\ell=1$, we can correspondingly yield $e^{\mathrm{i}\Delta\ell\theta}=\mathrm{i}$, attaining those as $v_p^{\Delta\ell=1}/DD(\rho_j)  =\sum_j2\kappa\{\delta_p\gamma_e^2\mathit{\Omega}^2(\rho_j)\mathcal{M}+g(\rho_j)\Omega_{\alpha}(\rho_j)\Omega_{\beta}^{*}(\rho_j)\Omega_{\zeta}(\rho_j)[\gamma_{e}\mathit{\Omega}^2(\rho_j)\mathcal{N}-g^{2}(\rho_j)\Omega_{\beta}^2(\rho_j)\mathcal{G}]\}$, and $ u_p^{\Delta\ell=1}/DD(\rho_j)  =\sum_j2\kappa\gamma_e\mathit{\Omega}^2(\rho_j)[\gamma_e\mathit{\Omega}^2(\rho_j)\mathcal{N}-g^{ 2}(\rho_j)\Omega_{\beta}^2(\rho_j)\mathcal{G}-g(\rho_j)\delta_p\mathcal{M}\Omega_{\alpha}(\rho_j)\Omega_{\beta}^{*}(\rho_j)\Omega_{\zeta}(\rho_j)]$.

Obviously, according to Eq. \eqref{Eq_0} and Eq. \eqref{Eq_2}, the conversion of the vortex phase factor between 0 and 2 will change the sign of $\gamma_e\mathit{\Omega}^2\mp g\Omega_{\alpha}\Omega_{\beta}^{*}\Omega_{\zeta}$ for single atom. In other words, the sign of the real and imaginary parts of the output field will be directly changed, which leads to distinct absorption/gain ($v_p$) phenomena. 

\section{The group delay of the output field}
Here we will use analytic calculations to show the form of the group delay. According to the phase of the relevant output field
\begin{align}
  \Theta=\arg[t_p],
\end{align}
we can get the conversion relationship as follows
\begin{align}\label{conversion}
  \frac{\partial\tan[\Theta]}{\partial\omega}=\frac{\partial\tan[\Theta]}{\partial\Theta}\cdot\frac{\partial\Theta}{\partial\omega}.
\end{align}
Since the group delay can be written as
\begin{align}
  \tau_g=\frac{\partial\Theta}{\partial\omega},
\end{align}
we can simplify it by using Eq. \eqref{conversion}, then the group delay has the form
\begin{align}
  \tau_g&=\frac{\partial\Theta}{\partial\omega}=\frac{\partial\tan[\Theta]}{\partial\omega}\cdot\frac{1}{1+\tan[\Theta]^2}\nonumber\\
  &=\frac{\partial\frac{v_p}{u_p-1}}{\partial\omega}\frac{(u_p-1)^2}{(u_p-1)^2+v_p^2}
\end{align}
where 
\begin{align}
  \tan[\Theta]=\frac{\mathrm{Im}[t_p]}{\mathrm{Re}[t_p]}=\frac{v_p}{u_p-1}.
\end{align}

\bibliography{Manu}

\end{document}